\documentclass[12pt]{article}

\usepackage{amsmath,amsfonts,epsfig,color,latexsym}

\newcommand{\be}{\begin{equation}}
\newcommand{\ee}{\end{equation}}
\newcommand{\bea}{\begin{eqnarray}}
\newcommand{\eea}{\end{eqnarray}}

\let\newsection=\section
\renewcommand{\section}{\setcounter{equation}{0}\newsection}

\begin{document}

\begin{flushright}
hep-th/0512171\\
BROWN-HET-1458
\end{flushright}
\vskip.5in

\begin{center}

{\LARGE\bf  DBI skyrmion, high energy (large s) scattering and 
fireball production}
\vskip 1in
\centerline{\Large Horatiu Nastase}
\vskip .5in

\end{center}
\centerline{\large Brown University}
\centerline{\large Providence, RI, 02912, USA}

\vskip 1in

\begin{abstract}

{\large We analyze the high energy scattering of hadrons in QCD in an 
effective theory model inspired from a gravity dual description. The 
nucleons are skyrmion-like solutions of a DBI action, and boosted 
nucleons give pions field shockwaves necessary for the saturation of the 
Froissart bound. Nuclei are analogs of BIon crystals, with the DBI 
skyrmions forming a fluid with a fixed inter-nucleon distance. 
In shockwave collisions one creates scalar (pion field)
``fireballs'' with horizons of nonzero temperature, whose scaling 
with mass we calculated. They are analogous to the 
hydrodynamic ``dumb holes,'' and their thermal horizons are places 
where the pion field becomes apparently singular.  The information 
paradox becomes then a purely field theoretic phenomenon, not directly 
related to quantum gravity (except via AdS-CFT).  
}

\end{abstract}

\newpage

\section{Introduction}

In a series of papers \cite{kn,kntwo,knthree,nastase,nastase2,nastase3}
 it was shown that the high energy, small angle
 (large s, fixed t) scattering in QCD can be described through a simple 
cut-off AdS gravity dual a la Polchinski-Strassler \cite{ps}, 
where the AdS space ends at an IR brane. Above the gauge 
theory Planck scale $\hat{M}_P=N_c^{1/4} \Lambda_{QCD}$, the scattering 
is dominated by black hole creation in the gravity dual. The maximal Froissart 
regime in QCD
\be
\sigma=\frac{\pi}{m^2}ln^2 s
\ee
corresponds in the gravity dual to black hole creation that occurs effectively 
on the IR brane. Then in the gravity dual we have an effectively 4d scattering 
in a gravity theory with a mass (the KK mass obtained by reducing gravity 
onto the IR brane), reducing to gravitational shockwave scattering in the 
large s limit, and creating black holes. It was shown in \cite{knthree}
 that this 
model exactly matches the 1952 Heisenberg model for the saturation of the 
Froissart bound \cite{heis}, a model in which one analyzes pion field 
shockwave 
scattering. Moreover, one needs a nonlinear DBI-type action for the pion 
field to obtain saturation of the bound. Heisenberg takes the action
\be
S=T_{(3)}
\int d^4 x \sqrt{1+\frac{(\partial_{\mu} \phi)^2}{\Lambda^2}+m^2\phi^2}
\label{heisdbi}
\ee
Because of the matching of the two descriptions of the Froissart saturation, 
in \cite{nastase3} 
 it was proposed that there should be a pion field ``soliton'' created
in the QCD collisions, radiating thermally at a given temperature, and this 
object was identified with the fireball observed at RHIC (see also 
\cite{sz,ssz} for a different view of how the RHIC fireballs are related 
to black holes, in ${\cal N}=4$ SYM). 

A different use of the pion field was put forward in the program of Skyrme-like
models \cite{skyrme,witten,rajeev,anw}. 
There one tries to find the nucleon as a topological solution of an 
effective pion field action with higher derivatives.  The nonlinear (or linear)
sigma model action does not admit solitons, but if one adds a higher order 
correction solitons can be obtained \cite{skyrme,rajeev,witten}. 
The original Skyrme model had a 
particular correction, but one obtains the same results with a large class 
of corrections. In particular, Pavlovskii \cite{pav} has analyzed a DBI-like 
action, that reduces to the sigma model at low energies and contains an 
infinite series of higher order corrections, summarized in the DBI square root.
The action 
\be
S=\frac{1}{2}
\int d^4 x f_{\pi}^2 \Lambda_{QCD}^2[\sqrt{1+(e^{-i\vec{\phi}_{\pi}\vec{
\tau}/f_{\pi}}\partial_{\mu}e^{i\vec{\phi}_{\pi}\vec{\tau}/f_{\pi}})^2
/(\Lambda_{QCD})^2}-1]
\ee
admits Skyrme-like topological solitons that could be identified 
with the nucleons. 

In this paper we will try to put these two approaches together, and find an
effective field theory description that will encompass both the hadrons at 
rest -nucleons in a Skyrme-like model- and the high energy scattering of 
hadrons. For the gravity dual description of the latter we often found 
direct effective field theory interpretation, so now we would like to 
use the intuition gained from the gravity dual to set up a purely effective 
field theory description of the scattering. We know that QCD itself should 
give a good description of the scattering, but a first principle QCD 
description is very hard, and the gravity dual description is suited to 
an effective field theory description anyway, so that's what we will find. 
However, this will not be a usual effective field theory, since if the 
Skyrme-like soliton only needs a few correction terms to the action, the 
large s scattering needs an action valid at energies well above the natural
cut-off, $\Lambda_{QCD}$.  

We will find an action that admits both a Skyrme-like topological 
soliton of the type of Pavlovskii and solutions with horizons that radiate 
thermally at a given temperature, corresponding to the ``fireballs'' observed
at RHIC. The fireball solutions will be very similar to the ``dumb holes'' 
of Unruh \cite{unruh}, 
obtained in sonic booms. The hydrodynamic equations of the sonic 
boom are exactly analogous to the scalar equations of motion in our theory.  

The paper is organized as follows. In section 2 we will review the high energy 
(large s) scattering in gravity duals of QCD, in section 3 we will look at 
various DBI actions and their BIon solutions. In section 4 we will boost the 
BIon solutions and compare to the boosting in the gravity dual model for 
QCD scattering. In section 5 we will analyze fireball-like BIon solutions 
with horizons and compare them with the ``dumb holes'', thus calculating 
their temperature. In section 6 we 
will look at SU(2) actions and topological solutions, in particular the 
Pavlovskii solution. Section 7 contains the bottom line of the paper.
We present our proposed DBI action, argue
for its form from the gravity dual perspective  together with 
QCD arguments and find its ``SkyrBIon'' 
solution. In section 8 we look at high energy scattering in our model and 
in section 9 we conclude.

\section{High energy scattering in QCD gravity duals}

In this section we review the description of large s, fixed t scattering 
via gravity duals of QCD.

Following Polchinski and Strassler \cite{ps}, the amplitude for
 scattering in QCD can be 
found from a gravity dual by multiplying with wave functions in the extra 
dimensions and integrating over these extra dimensions
\be
{\cal A}(p)=\int dr d^5\Omega \sqrt{g}{\cal A}_{string}(\tilde{p}) \prod_i
\psi_i
\label{psrel}
\ee
We do not know of course the gravity dual of QCD, but we know that it should 
look like $AdS_5\times X_5$ space modified in the IR (and maybe in the UV).
\be
ds^2= \frac{\bar{r}^2}{R^2} d\vec{x}^2 + \frac{R^2}{\bar{r}^2}d\bar{r}^2 + 
R^2 ds_X^2= e^{-2y/R}d\vec{x}^2 + dy^2 + R^2 ds_X^2
\ee
Polchinski and Strassler proposed that one could obtain a lot of information 
just by putting a sharp cut-off in the IR (at $r_{min}\sim R^2 \Lambda$).
In the large s, fixed t regime, the scattering in the gravity dual
is concentrated close to, but not on the IR cut-off (IR brane) \cite{kntwo}. 

But high s, fixed t scattering in a gravitational theory was shown by 
't Hooft \cite{thooft} to be described by scattering Aichelburg-Sexl  
gravitational shockwaves \cite{as}
\be
ds^2 = 2dx^+ dx^- +(dx^+)^2 \Phi (x^i) \delta (x^+) +d\vec{x}^2
\label{shock}
\ee
where the function $\Phi $ satisfies the Poisson equation
\be
\Delta_{D-2} \Phi (x^i) = -16 \pi G p \delta ^{D-2}(x^i)
\label{poisson}
\ee
He showed that for $s\leq M_{Pl}$ one 
can treat one particle as a shockwave and the other as a null probe (geodesic),
and proposed that above $M_{Pl}$ both particles should be shockwaves, and 
one will create black holes. This case of $s\gg M_{Pl}$
was analyzed (and the cross section for black hole production was 
computed) in flat 4 dimensions in \cite{eg} and extended to higher 
dimensions and curved space in \cite{kn}. As suggested in \cite{gid} on 
general arguments, it was found that the cross section in flat space grows 
like a power law 
\be
\sigma\sim r_H^2\sim s^{\frac{1}{D-3}}
\ee

One can find solutions for gravitational shockwaves in curved spaces of 
gravity dual type \cite{nastase}, and we find that the solutions still 
look like (\ref{shock}) in the given background, where now $\Phi$ satisfies 
the Poisson equation in the background. 

In \cite{kntwo,nastase2} it was found that the behaviour for the 
cross section for black hole formation in the gravity dual translates into 
the same kind of behaviour for the QCD cross section, and we have the 
following regimes. 

Above the Planck scale, which in gauge theories corresponds 
to $\hat{M}_P=N_c^{1/4}\Lambda_{QCD}$ with $\Lambda_{QCD}$ the scale of the 
lightest glueball excitation, and in real QCD would be about 1-2 GeV, in 
the gravity dual we form small black holes, that ``feel'' only flat space. 
Correspondingly, one finds that the shockwave profile $\Phi$ is 
\be
\Phi = \frac{16\pi G_D}{\Omega_{D-3}(D-4) r^{D-4}}\sim \frac{1}{r^{D-4}}
\ee
By applying the formalism for shockwave scattering with black hole formation 
and translating to QCD with (\ref{psrel}) one finds that 
\be
\sigma \sim s^{\frac{1}{D-3}}
\ee
both in the gravity dual and in QCD. 

As one increases s beyond $E_R=N_c^2/R$ in the gravity dual and above 
$\hat{E}_R=N_c^2\Lambda_{QCD}$ in the gauge theory, about 10 GeV in real 
QCD, the black holes produced in the gravity dual start to ``feel'' 
the curvature
of space. One finds that for consistency, in the $AdS_{d+1}\times X_{\bar{d}}$
gravity dual $X_{\bar{d}}$ needs to be large (with scale much larger in the IR
than that of AdS), and the shape of the shockwave profile becomes 
\be
\Phi=\frac{K_1 R_s R^n}{r^n}\sim \frac{1}{r^{2(d-1)+\bar{d}}}= \frac{1}{r^{11}}
\ee
and by applying the formalism one finds 
\be
\sigma\sim s^{1/n}=s^{1/11}
\ee
both in the gravity dual and in QCD.

Finally, the last regime corresponds to the maximal Froissart behaviour. 
Above an unknown energy scale $\hat{E}_F$, that depends on the details of the 
gravity dual in the IR, but in real QCD the experimental data suggests it 
should be between 100 GeV and 1 TeV, the black holes created in the gravity 
dual are so large that they reach the IR cut-off and get stuck there (since 
the scattering in the dual happens mostly near the IR). Thus the gravity 
dual scattering effectively happens on the 4d IR brane, and creates 4d black 
holes. The shockwave profile is 
\be
\Phi(r, y=0)\simeq R_s \sqrt{\frac{2\pi R}{r}}C_1 e^{-M_1 r}
\sim  e^{-M_1 r}
\ee
where $M_1$ is the mass of the lightest graviton excitation: if we are reducing
gravity on the IR brane, gravity has a nonzero KK mass $M_1$. Then one 
finds the gauge theory cross section
\be
\sigma_{gauge}\simeq \bar{K}\pi [\frac{\sqrt{2}}{M_1}ln 
[0.5 \sqrt{s}M_1\hat{G}_4]]^2
\ee
where $\bar{K}$ is a numerical constant depending on the details of the gravity
dual, $M_1$ translates to the mass of the lightest QCD excitation and 
the constant multiplying $\sqrt{s}$ in the log cannot be taken too seriously, 
as the subleading behaviour of $\sigma$ is modified anyway. 

The description of 
this last Froissart behaviour was shown in \cite{knthree} to match exactly to 
the description of the saturation in Heisenberg's model \cite{heis}. 
Heisenberg says that at high enough energies the Lorentz-contracted hadrons 
colliding will look like shockwaves, characterized by a transverse size 
$\sim 1/M_H$. Moreover, at high enough energies, the hadrons effectively 
``dissolve'' in the pion field, which also becomes Lorentz contracted to 
a shockwave, with transverse size characterized by $1/m_{\pi}$. 

Heisenberg assumes that the ``degree of inelasticity'' $\alpha$
(=${\cal E}/ \sqrt{s}$= energy loss/collision energy) behaves like the 
overlap of pion wavefunctions, and the pion wavefunction behaves like 
$\psi (x^i)\sim e^{-m_{\pi} r}$. Then one finds that the behaviour of the 
QCD cross section is given by
\be
\sigma =\pi b_{max}(s)^2 \simeq \frac{\pi}{m_{\pi}^2}ln^2 \frac{\sqrt{s}}{
<E_0>}
\ee
where $<E_0>$ is the average emitted pion energy. But the average emitted 
pion energy is found to increase linearly with energy for a free pion action 
or for an action of the $\lambda \phi^4$ type. Instead, Heisenberg finds that 
one needs a nonlinear action in order 
to get an approximately constant $<E_0>$. 
He takes the DBI action in (\ref{heisdbi}) and then he gets 
$<E_0>\sim m_{\pi}$. 

The same kind of picture appears in the gravity dual, where we collide 
particles, characterized by some size, but in the high energy limit 
we find that the only thing of relevance is the gravity field, and we 
effectively collide gravitational shockwaves, also characterized by a shockwave
profile $\Phi$ that has a characteristic size $M_1$, the scale of the 
lightest excitation.

Until now, we talked in the gravity dual about pure gravity, corresponding 
to pure gauge theory. But in reality QCD has pions (Goldstone bosons of 
chiral symmetry) which are much lighter than the scale of the lightest 
glueball. The discussion of the gravity dual before the onset of the Froissart
behaviour is the same, as it is governed still by gravity producing black 
holes.  But if one has a Goldstone boson in the gauge theory, it can be 
modelled by a radion in the gravity dual (the position of the IR brane
can be made dynamical). If the radion mass is smaller than $M_1$ (the KK 
graviton mass), the radion will dominate at large enough s (dominated by 
the IR). The scattering will then produce local brane bending, and the 
brane will bend significantly, entering the effective scattering region 
of the gravity dual before the black holes created in this region 
will become large enough, and thus the cross section at infinite s
will be dominated  by brane bending.  
If the radion is more massive than gravity ($M_1$), 
the brane bending will be small, and the created black holes will reach the 
brane before the brane reaches the scattering region, and the cross section 
at infinite s will still be dominated by black holes.

In the physical case of lighter radion (thus lighter pion), at least
 heuristically the same Froissart behaviour will apply, as shown in \cite{gid}.
We do not have at our disposal the rigorous arguments of \cite{kntwo} anymore,
since those were based on powerful general relativity
 theorems about horizons, and the 
collision of shockwaves in scalar field theory is still too hard to 
solve explicitly, but we will still  rely on our gravity intuition for 
guidance. 

One observes that the action for the radion, considered as a brane moving in 
one (almost flat) direction, is in fact the DBI action
\be
S=T_{(3)}\int d^4 x \sqrt{1+(\partial_{\mu}X)^2/\Lambda^2}
\ee
where $X$ is the radion (brane position) and $\Lambda$ is related 
to the string scale.  
This is exactly the action taken by Heisenberg (\ref{heisdbi}), at zero mass, 
but one could consider as a radion stabilization mechanism giving it a mass 
according to (\ref{heisdbi}). Moreover, if the IR brane is considered to be 
a D-brane probe, it will also have a U(1) gauge field on it. On static 
solutions (time independent) and at zero magnetic field, the action can 
be taken as 
\be
S= T_{(3)}
\int d^4 x \sqrt{1+(\nabla X)^2/\Lambda_1^2-(\nabla \phi)^2/\Lambda_2^2}
\ee
where $\phi$ is an electric potential and $\Lambda_1$ and $\Lambda_2$
can be a priori different. We have to remember though that this is the 
D-brane action in a flat extra dimension, thus the curvature of the gravity 
dual space will induce higher order corrections to the DBI action. We will 
discuss them later. 

We are thus driven to study the DBI actions with both signs inside the 
square root, corresponding to either real scalar or electric potential 
(coming from the original Born-Infeld action).

\section{DBI actions and BIon solutions}

In this section we look at DBI actions and their BIon solutions and try to 
connect to the picture from the last section. 

We saw that both in the Heisenberg model and in the gravity dual description 
we are driven to consider DBI actions. In the gravity dual shockwave 
collisions produced black holes, and equivalently we proposed that 
in QCD pion field shockwave collisions should produce ``fireballs'', 
solutions with horizons radiating thermally. The collision process is 
quite complicated, as we saw in the gravity dual case, where we could 
prove a black hole forms, but we couldn't calculate its metric, so we 
took as an approximation that we create a static spherically symmetric 
black hole. In reality however we have a dynamic complicated process.
Similarly now we will take as an approximation that we create static 
spherically symmetric ``fireballs,'' thus in this section we will study 
such solutions.

Moreover, one can ask also whether one can understand the nucleons as being 
modelled by solutions to the same DBI actions. We will thus 
look at all static spherically symmetric 
solutions of the DBI actions. In this section we will set the DBI 
scales $\Lambda_1$ and $\Lambda_2$ to 1 for simplicity.

{\bf DBI action for 4d YM}. The original action of Born and Infeld \cite{bi}
is
\be
{\cal L}= \sqrt{-det (\eta_{\mu\nu}+\frac{F_{\mu\nu}}{\sqrt{2}})}=
\sqrt{1+\frac{F_{\mu\nu} F^{\mu\nu}}{2}-(\frac{F_{\mu\nu}*F^{\mu\nu}}{2})^2}=
\sqrt{1-\vec{E}^2 +\vec{B}^2-(\vec{E}\cdot \vec{B})^2}
\ee
Here $E_i=F_{0i}, F_{ij}=\epsilon_{ijk} B_k$. One defines in the usual way 
\be
\vec{D}=\frac{\partial {\cal L}}{\partial \vec{E}}
\ee
Then at B=0, the field equation is 
\be
\nabla \cdot \vec{D}=\rho
\label{max}
\ee
as in electromagnetism (Maxwell), just that now we have a different definition
for $\vec{D}$
\be
\vec{D}=-\frac{\vec{E}}{\sqrt{1-\vec{E}^2}}\Rightarrow \vec{E}=
-\frac{\vec{D}}{\sqrt{1+\vec{D}^2}}
\ee

Defining the electric potential $\phi$ by $\vec{E}= -\nabla \phi$
we get the DBI-electric lagrangean
\be
{\cal L}=\sqrt{1-(\nabla \phi)^2}
\ee

{\bf 4d scalar DBI}. We follow the above analysis closely.
In the static case ($\partial _t =0$), we define
\be
\vec{F}= \vec{\nabla} X
\ee
The Lagrangean is then 
\be
{\cal L}= \sqrt{-\det(\eta_{\mu\nu}+\partial_{\mu}X\partial_{\nu}X)}
=\sqrt{1+(\partial_{\mu}X)^2}= \sqrt{1+\vec{F}^2}
\ee
Defining 
\be
\vec{C}= \frac{\partial {\cal L}}{\partial \vec{F}}
\ee
the equation of motion is
\be
\nabla \cdot \vec{C}= \rho 
\ee
as for the free theory! (analogous to the fact that for BI we had the same
equation as for the Maxwell case) except now the definition of $\vec{C}$ 
is different
\be
\vec{C}= \frac{\vec{F}}{\sqrt{1+\vec{F}^2}}\Rightarrow
\vec{F}= \frac{\vec{C}}{\sqrt{1-\vec{C}^2}}
\ee

{\bf BIons and catenoids}

For a treatment of solutions to the scalar and electric DBI action
see \cite{cm, gibbons, review}. The electric BIon (solution to electric 
DBI), originally found by Born and Infeld, is given by 
\be
\phi(r)= C\int_r^{\infty} \frac{dx}{\sqrt{C^2+x^4}}
\ee
with the asymptotics (eqs 72-75 in \cite{gibbons})
\be
\phi(r)\simeq \frac{C}{r}, \;\; r\rightarrow \infty;\;\;\;\;
\phi(r)\simeq const. -r , \;\; r\rightarrow 0
\ee
From the asymptotics at infinity, we see that $C=q$=electric charge. 
The BIon is found by noting that the solution to (\ref{max}) 
with delta function source is the same as 
for the Maxwell theory in terms of $\vec{D}$, and then finding $\phi$. 
As Born and Infeld noted, the advantage is now that the solution is nonsingular
in terms of $\vec{E}$, which reaches its maximum of $\vec{E}=1$ at r=0. 
Consequently, also the energy of this solution (with delta function source) 
is finite, unlike the case of Maxwell theory. 

Analogously, one finds the ``catenoid'' solution, i.e. solution to the 
scalar DBI action, as 
\be
X(r) =\bar{C}\int_r^{\infty} \frac{dx}{\sqrt{x^4-\bar{C}^2}}
\ee
But this has a horizon at $r=r_0=\sqrt{C}$, 
where $\vec{F}=\vec{\nabla} X $ diverges, even 
though $X$ remains finite. 

{\bf Solutions to the D-brane action}

As shown in \cite{cm}, one can embed the previous solutions into the 
U(1) D-brane action, that on static solutions (and at zero magnetic field) is 
\be
S=T_{(3)}
\int d^4 x \sqrt{ (1-(\nabla \phi)^2)(1 +(\nabla X)^2)+(\vec{\nabla}\phi
\cdot \vec{\nabla}X)^2}
\label{daction}
\ee
For a review of solutions to this action, see also \cite{review,gibbons}.
On top of those, the D-brane action also has BPS solutions, which were shown 
to correspond to ($C=\bar{C}\equiv q$)
\be
\phi=X= \frac{q}{r}
\ee
and are understood in string theory as fundamental strings attached to the 
D brane and stretching all the way to infinity (BPS BIons). It has a 
singularity at r=0, where the fundamental string is attached, and has 
infinite energy, because the fundamental string has finite tension and 
infinite length. 

The BIon solution however has no singularity nor a horizon at $r=\sqrt{C}$.
 One solves the 
equation for $\vec{D}$ in the usual way ($\vec{D}=\vec{\nabla} (C/r)$), 
and at $r=\sqrt{C}$, $|\vec{D}|=1$ and $|\vec{E}|=1/\sqrt{2}$, and there is no 
singularity: the electric potential $\phi$ is continuous and its derivative
is also. The BIon has a finite total energy, which is the reason Born and 
Infeld put it forward in the first place, as they wanted to have an electron 
(solution with delta function source) with finite classical self-energy. 
In string theory it is not clear what its interpretation is, but one might 
not worry about that too much, since it carries no topological charge,  so 
one could maybe have doubts about its stability. The simplest possibility 
for an interpretation is 
of a string going {\em through} the D-brane, such as not to excite the scalar 
field. 

An interesting property of BIons is that like charges
 repel each other, and opposite charges attract (just as in Maxwell theory), 
and one is able to write down explicitly a BIonic crystal solution 
(first done by Hoppe \cite{hoppe}, see also \cite{gibbons}), in which each 
charge is surrounded by opposite charges, like in a NaCl crystal of Maxwell 
theory. 

For the ``catenoid'' solution, at $r=\sqrt{C}$, 
$X$ is finite but not its derivative, and we will call this surface a 
horizon. We will note later on that this name is justified, 
and it is very like the horizon of a black hole, but for 
the moment it just indicates the aparent singularity of the solution.
Putting back momentarily an energy scale $\beta^2$, i.e. from 
\be
S=\beta^{-2}\int [\sqrt{1+\beta^2 (\partial X)^2}-1]+ \int X (\bar{C}
\delta (r))
\ee
we have
\be
C_i=\partial_i \frac{\bar{C}}{r};\;\;\; F_i =\frac{C_i}{\sqrt{1-\beta^2 \vec{C}
^2}}
\ee
We can ask the question: can we continue the solution beyond this 
aparent singularity?

\begin{figure}

\begin{center}

\includegraphics{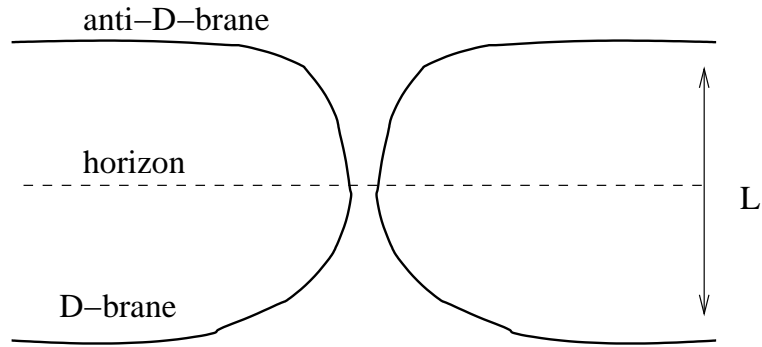}
\end{center}
\caption{D-brane anti-D-brane system connected by a string: two catenoid 
solutions with different asymptotic regions connected on the horizon}
\label{ddbar}
\end{figure}

Unlike for a black hole, it was argued in \cite{cm}
that in this case the only analytic 
continuation that makes sense is to glue another solution with a
different asymptotic space, creating an Einstein-Rosen bridge.
In string theory, this corresponds to a brane-
antibrane pair, connected by a fundamental string of length L, 
as in fig.\ref{ddbar}.

\begin{figure}

\begin{center}

\includegraphics{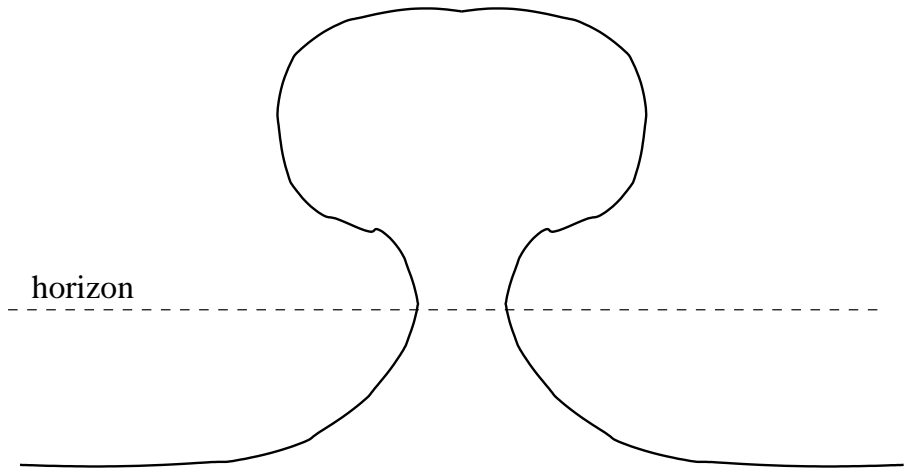}
\end{center}
\caption{A single D-brane, the continuation past the horizon can be 
a brane ``bubble''}
\label{cont}
\end{figure}

But if we think of the catenoid as being a metastable solution, created in 
a collision of some type, as we will want later on, one should not have a 
full extra anti-D-brane, but one should at least be able to continue 
into something that looks {\em locally } as an extra brane (a brane 
``bubble'') as in fig.\ref{cont}, if not in a different way.


For a black hole, when we reach the horizon, we can analytically continue 
in two ways: we can either continue to a singularity, OR glue another 
asymptotic region to create an Einstein-Rosen bridge. So if the catenoid 
is a scalar analog of a black hole, we would think we should be able to 
continue the solution to r=0 without making an Einstein-Rosen bridge.

However if we naively continue $\vec{F}$ through $r=\sqrt{\bar{C}}$ 
using the same 
relation relating it to $\vec{C}$ (where $\vec{C}$ is the solution to 
$\nabla \cdot \vec{C}=\bar{C}\delta (\vec{r})$), it becomes complex, thus 
$X$ becomes complex. So the only possibility of keeping $\vec{F}$ and 
$X$ real is to change the sign of $(\partial X)^2$ 
in the action, effectively changing 
$X$ into a $\phi$ (electric potential)-type variable. 
Then the continuation 
is analytic in the sense that $\vec{C}$ has the same (analytic) expression 
inside and outside $r=\sqrt{\bar{C}}$
 (for it, nothing interesting happens at $r=\sqrt{\bar{C}}$). 
But it is not clear what would be the physical significance of changing the 
sign of $(\partial X)^2$ in the action at the horizon. 

Note that
\be
\phi(\sqrt{C})=C\int_{\sqrt{C}}^{\infty} \frac{dx}{\sqrt{x^4+C^2}}= 
\sqrt{C}\frac{2\Gamma(5/4)^2}{\sqrt{\pi}}= C\int_0^{\sqrt{C}}
 \frac{dx}{\sqrt{x^4+C^2}}
\ee
thus $\phi(0)= 2\phi(\sqrt{C})$, whereas
\be
X(\sqrt{\bar{C}})= \bar{C}\int_{\sqrt{\bar{C}}}^{\infty}\frac{dx}{\sqrt{x^4-
\bar{C}^2}}= \sqrt{\bar{C}}\sqrt{\pi} \frac{\Gamma(5/4)}
{\Gamma (3/4)}=\sqrt{2} \phi (\sqrt{C})|_{C=\bar{C}}
\ee
so gluing the $X$ solution to a $\phi$-type
 solution inside the horizon would 
give a finite displacement $X(0)$ at r=0 ($X$ is the position of the
D-brane in the transverse direction). 

For the BIon action $\sqrt{1-(\vec{\nabla} {\phi})^2}$, 
we have also the solution
\be
\phi (r) = C\int_r^{\infty} \frac{dx}{\sqrt{C^2-x^4}}
\ee
which is just the continuation of the ``catenoid'' solution to $r<r_0$, and 
redefining $X \rightarrow iX$, thus effectively changing the action.
This solution also has a horizon at $r=r_0$, where $\phi$ is finite, but 
$\phi '$ is infinite and negative. However, this solution has imaginary 
action (or energy) (the square root is negative), but is a real slution 
to the equations of motion.

Finally, note that the BIon and catenoid solutions are defined up to a 
sign ($\pm$), and an additive constant. One could construct more solution 
that are not real. For instance, $i\times$ the BIon solution is a solution 
to the scalar (``catenoid'') action, but that is a trivial observation, 
reflecting the fact that redefining the scalar by an i takes us from one 
action to the other.

For a general static solution of the D-brane action studied in \cite{cm}, the 
equations of motion can be written as 
\bea
&&  \vec{\nabla} \cdot g_p \vec{\Pi}=  \vec{\nabla} \cdot[
\frac{\vec{E}(1+(\vec{\nabla}X)^2)
-\vec{\nabla}X(\vec{E}\cdot \vec{\nabla}X
)}{\sqrt{(1-\vec{E}^2)(1+(\vec{\nabla}X)^2)
+ (\vec{E}\cdot \vec{\nabla}X)^2}}]=0\nonumber\\
&&  \vec{\nabla} \cdot[\frac{\vec{\nabla}X+g_p^2\vec{\Pi} (\vec{\Pi}\cdot \vec{
\nabla}X)}{\sqrt{1+(\vec{\nabla}X)^2+g_p^2\vec{\Pi}^2 +
g_p^2(\vec{\Pi}\cdot \vec{\nabla}
X)^2}}]=0 \; or\; \nonumber\\&&
\vec{\nabla} \cdot[\frac{\vec{\nabla}X(1-\vec{E}^2)+\vec{E} (\vec{E}\cdot
\vec{\nabla}X)}{\sqrt{(1-\vec{E}^2)(1+(\vec{\nabla}X)^2)
+ (\vec{E}\cdot \vec{\nabla}X)^2}}=0
\eea

On  a spherically symmetric solution (with $\vec{E}=-\vec{\nabla}\phi$), 
the equations become
\be
[\frac{r^2 \phi '}{\sqrt{1-\phi'^2 +X'^2}}]'=0;\;\;\;
[\frac{r^2 X '}{\sqrt{1-\phi'^2 +X'^2}}]'=0
\ee

The solutions of these equations are obtained by integrating them with 
constants C and $\bar{C}$, respectively:
\be
{\phi '}^2 =\frac{1+{X'}^2}{1+r^4/C^2};\;\;\;\;
{X'}^2=\frac{1-{\phi '}^2}{r^4/\bar{C}^2-1}
\ee
from which we obtain the general solution
\bea
&&X'=\frac{\bar{C}}{\sqrt{r^4+C^2-\bar{C}^2}}\nonumber\\
&&\phi '=\frac{C}{\sqrt{r^4+C^2-\bar{C}^2}}
\eea
Thus we see that for $C>\bar{C}$, both $\phi$ and X look like the BIon, 
being everywhere defined and having finite derivatives at zero. For 
$C<\bar{C}$, both $\phi$ and X look like the catenoid, having a horizon 
at a finite r, equal to $(\bar{C}^2-C^2)^{1/4}$. At $C=\bar{C}\equiv q$ 
we have the BPS BIon, which blows up at r=0. 

At r=0, we can specify $\bar{\phi}(0)\equiv \phi '(r=0)$ and 
$\bar{X}(0)\equiv X'(r=0)$, but we have to satisfy 
\be
(1-\bar{\phi}(0)^2 +\bar{X}(0)^2)\bar{\phi}(0)=
(1-\bar{\phi}(0)^2 +\bar{X}(0)^2)\bar{X}(0)=0
\ee
i.e. either $\bar{\phi}(0)=\bar{X}(0)=0$, or 
$\bar{\phi}(0)^2-\bar{X}(0)^2=1$ (for the BPS BIon $ \bar{\phi}(0)=\bar{X}
(0)=\infty$, so that solves it). 
We can check explicitly from the general solution
that this equation is true. 
The BIon has $\bar{\phi}(0)=1$, thus we can 
think of any solution with X as a perturbation around the BIon. 
Note that this means that the square root in 
the action is equal to zero at $r=0$. Only for the BPS BIon the square root 
at $r=0$ is equal to 1. 

At $r=\infty$, by putting 
\be
\bar{\phi}\sim \frac{a}{r^n}\;\;\; \bar{X}\sim \frac{b}{r^m}
\ee
we find from the equations of motion that $n=m=2$, as for the separate 
$\phi$ and $X$ theories, thus the interaction doesn't change that, as it 
should. 

As we saw, if we put $\bar{\phi}=0$, any solution will have a horizon
(the unique solution is of catenoid type, which has a horizon). 
We check that indeed, we can't have a real solution defined at r=0, 
since then $-\bar{X}(0)^2=1$. However, 
if $\bar{\phi}\neq 0$, we see that the solution has a finite $\bar{X}(0)
=\sqrt{\bar{\phi}(0)^2-1}$, and the solution can be extended to infinity
without encountering a horizon. This is due to the fact that we can think 
of these solutions as adding more electric field at r=0 ($\bar{\phi}(0)-1$)
to the BIon, in order to compensate for the added scalar ($\bar{X}(0)$).  

The Hamiltionian of a static configuration ($\dot{X}=0$) is 
\be
H=\frac{1}{g_p}\int d^px[\frac{1+(\vec{\nabla} X)^2}{\sqrt{(1-\vec{E}^2)
( 1+(\vec{\nabla} X)^2)+(\vec{E}\cdot \vec{\nabla}X)^2}}-1]
\ee
For the general spherically symmetric 4d static configurations above it is 
\be
E=\frac{1}{g}\int d^3 x  [\frac{1+{X'}^2}{\sqrt{1-{\phi '}^2+{X'}^2}}-1]
=\frac{1}{g}\int d^3 x [\frac{r^4+C^2}{\sqrt{r^4(r^4+C^2-\bar{C}^2)}}-1]
\ee
Specifically, for the catenoid (purely scalar), 
\be
gE_X= \int _{r_0}^{\infty}r^2 dr (\sqrt{1+{X'}^2}-1)= 
\bar{C}^{3/2}I
\ee
where 
\be
I=\int_1^{\infty}x^2 dx [\frac{x^2}{\sqrt{x^4-1}}-1]\simeq 0.770343
\ee
Note that then $r_0=\sqrt{\bar{C}}\propto (E_X)^{1/3}$, same as for an 
object of constant energy density. However, the energy density diverges 
near the horizon, showing that a significant fraction of the energy is 
concentrated near it. For the BIon we have 
\be
g E_{\phi}=\int_0^{\infty} r^2 dr [\frac{1}{\sqrt{1-{\phi '}^2}}-1]
=C^{3/2}\int_0^{\infty}\frac{dx}{x^2+\sqrt{x^4+1}}=\frac{(\Gamma[1/4])^2}{6
\sqrt{\pi}}C^{3/2}
\ee
and now the energy density diverges at r=0, thus a significant portion of the 
energy is situated near the origin. 
Then for the general solution with horizon ($\bar{C}>C$) we have 
\be
gE= r_0^3I+\frac{C^2}{r_0}\sqrt{\pi}\frac{\Gamma[5/4]}{\Gamma[3/4]};\;\;\;\;
r_0^4\equiv \bar{C}^2-C^2
\ee
and for the general solution with no horizon ($C>\bar{C}$)
\be
gE= \frac{(\Gamma[1/4])^2}{2\sqrt{\pi}\bar{r}_0}[\frac{\bar{r}_0^4}{3}+
\frac{\bar{C}^2}{2}]
;\;\;\;\; \bar{r}_0^4\equiv C^2-\bar{C}^2
\ee
For the BIon C is an asymptotic U(1) charge, thus will be quantized in 
the quantum theory, and for the catenoid $\bar{C}$ is an asymptotic 
scalar charge, thus again we expect it to be quantized. We see that both 
the catenoid and the BIon energies go like charge to the power 3/2, thus 
higher charge objects will be unstable towards decay onto the lower charge 
ones ($E(Q_1+Q_2)>E(Q_1)+E(Q_2)$)
 This is also valid for the general solution with both $C$ and $\bar{C}$
nonzero (if $C\propto \bar{C}\propto n$, then $E\propto n^{3/2}$ always). 

Due to the nonlinearity of the action, it is hard to construct explicit 
(separated) multi center solutions. 
The classical interaction potential between 
two solutions would be then given by the difference between the energy 
of the two-center  solutions and the individual energies of the single center
solutions,
\be
E(R)= E(1, r_0=0; 2, r_0=R)-E_1-E_2
\ee
Given that we can't construct the multicenter solutions, we can't calculate 
the form of the potential, but we can say something about the asymptotic 
features. At large distances, the interaction becomes (free) 
Maxwell electromagnetism plus free scalar, thus the potential between two 
BIons will be 
\be
E_\phi(R)\simeq \frac{Q_1Q_2}{R}
\ee
where $Q_1$ and $Q_2$ are the electric charges ($C_1$ and $C_2$), thus 
repulsive if $Q_1Q_2>0$ and attractive if $Q_1Q_2<0$. And the 
potential between two catenoids will be 
\be
E_X(R)\simeq -\frac{\bar{Q}_1\bar{Q}_2}{R}
\ee
thus attractive for like scalar charges $\bar{Q}_1$ and $
\bar{Q}_2$ ($\bar{C}_1,\bar{C}_2$). For a general solution, it should be 
the sum of the two potentials. 

However, at small enough distances, the two center solution will look 
approximately like a single center solution with charges equal to the 
sum of the individual charges, thus the potential will always be repulsive!
That is true both for the BIons, and for the catenoids, thus for the general 
solution, since in all cases the energy goes like charge to the power 3/2. 
This is  of course for the case when the interacting objects are of the same 
type (same ratio $\bar{C}/C$).

The only exception is the BPS BIon, for which the energy of two center 
BPS objects exactly equals the sum of the individual BPS objects, i.e. 
the potential is zero. In that case, the potential is zero at infinity
($E_{\phi}$ cancels against $E_X$), but what about at zero? 

The energy of the BPS BIon is 
\be
gE= \int d^3 x (\vec{\nabla }X)^2=
\int_0^{\infty}r^2 dr [\frac{r^4+C^2}{r^4}-1]= C^2\int_0^{\infty}\frac{dr}{r^2
}
\ee
thus divergent, and the divergence is the same as in the free scalar theory
(or free electromagnetism). Because the energy is divergent, we can't 
draw any conclusions from the fact that it behaves like the square of the 
charge. In fact, the energy of the multicenter solution is exactly equal to 
the sum of the individual solutions, due to the BPS property. The point is
here that one has to regularize the infinite energy, and if one introduces 
a lower cut-off for r, $r\geq \delta$, while keeping also the value of 
the scalar field, $X(\delta)$, fixed, one gets 
\be
gE\sim \frac{C^2}{\delta}\propto CX(\delta)
\ee
thus energy that is 
linear in the charge C. The divergence signals the fact that we 
have to take into account quantum theory, and at the quantum level we know 
the BPS BIon is a BPS string, thus stable, and in string theory
$X(\delta)$ has physical significance. Thus the asumption is that $X(\delta)$
is kept fixed, i.e. $X_1(\delta)=X_2(\delta)=X_{1+2}(\delta)$, and one 
just adds the coefficient C. For a catenoid we have also 
$X(\bar{r}_0)=\bar{C}^{1/2}\sqrt{\pi}\Gamma[5/4]/\Gamma[3/4]\simeq 
1.31103 \bar{C}^{1/2}$, thus $E\propto \bar{C}X(\bar{r}_0)$, but now the 
energy is finite, and we can treat this object classically, and 
there is no reason to keep $X(\bar{r}_0)$ fixed, but we can let it vary with 
$\bar{r}_0$, and there is 
no interpretation in terms of a string ($\bar{r}_0$ 
is just a finite ``thickness'' of the solution). 

The same situation of diverging energy 
 would happen if we tried to apply the above logic to a free theory
like Maxwell electromagnetism or free scalar theory. For example, for 
electromagnetism, the classical potential would be 
\bea
&&E(R)= \int d^3 x [\vec{E}^2(1, r=0;2, r=R)-\vec{E}^2_1-\vec{E}^2_2]
\nonumber\\ &&
= \int r^2 dr [(\frac{Q_1\hat{r}}{r^2}+\frac{Q_2 (\vec{r}-
\vec{R})/|\vec{r}-\vec{R}|}
{(\vec{r}-\vec{R})^2})^2- (\frac{Q_1\hat{r}}{r^2})^2-
(\frac{Q_2 (\vec{r}-\vec{R})/|\vec{r}-\vec{R}|}{(\vec{r}-\vec{R})^2})^2]
\nonumber\\ &&
= 2 Q_1Q_2 \int r^2 dr \frac{\hat{r}\cdot (\vec{r}-\vec{R})/|\vec{r}-\vec{R}|}{
r^2(\vec{r}-\vec{R})^2}
\eea
and this diverges, and we can only draw the conclusion that 
$E(R)\sim Q_1Q_2/R$ by scaling, 
which is correct, but nothing about the coefficient 
or its sign. In fact exactly the same calculation applies for the free scalar, 
but for scalars the true sign is opposite!

\section{Boosting BIons; comparison with gravity}

In order to have a consistent effective field theory picture for high 
energy scattering, we should boost the solutions corresponding to the 
nucleons, creating shockwaves in the pion field. The collision of such 
shockwaves should create ``fireball''-like objects. In this section we 
will thus analyze the boosting of the DBI solutions from last section.
We will then compare with the gravity dual case.

{\bf Boosted ``catenoid''}

If one boosts a scalar static solution $f(x,y,z)$ , by definition the 
boosted solution is $g(x', y', z', t')= f(x, y, z)$, thus for a spherically 
symmetric solution $f(\sqrt{x^2+y^2+z^2})$ we have 
\be
g(x, y, z, t)= f(\sqrt{x^2+y^2+\frac{(z-vt)^2}{1-v^2}})
\ee
and it thus satisfies the equations
\be
\partial_{z'} g= \gamma \partial _z f|_{z=\gamma(z'-vt)}
 ;\;\; \partial_{y'} g= \partial _y f|;\;\;
\partial_{x'} g= \partial _x f|;\;\; \partial_{t'} g= -\gamma v \partial _z f|
\ee

Let's then boost the ``catenoid'', 
\be
X(r)= \int_r^{\infty} \frac{\bar{C} dx}{\sqrt{x^4-\beta^2 \bar{C}^2}}
\ee
This is a true scalar, so we just replace r with 
\be
r=\sqrt{\tilde{r}^2+\gamma^2 (z-vt)^2};\;\; \tilde{r}^2=x^2+y^2
\ee
Then after the boost
\bea
&& \frac{\partial X}{\partial \tilde{r}}= \frac{\bar{C}}{\sqrt{r^4-\beta^2 
\bar{C}^2}}
\frac{\tilde{r}}{r}; \;\;\;
 \frac{\partial X}{\partial z}= \frac{\bar{C}}{\sqrt{r^4-\beta^2 \bar{C}^2}}
\frac{\gamma^2 (z-vt)}{r};\nonumber\\
&& \frac{\partial X}{\partial t}= -\frac{\bar{C}}{\sqrt{r^4-\beta^2 \bar{C}^2}}
\frac{\gamma^2 v(z-vt)}{r}\;\;\;
 \Rightarrow (\partial X)^2 =\frac{\bar{C}^2}{r^4 -\beta^2 \bar{C}^2}\;\;\;
{\rm as \;\; before\;\; boost}
\eea
Thus in the limit $v\rightarrow 1$ we have the shockwave-like solution 
\bea
X&=& X (\tilde{r}), \;\;\; z=t\nonumber\\
&& 0;\;\;\; {\rm otherwise}
\eea
(one could write $X = X(\tilde{r}) \delta (z-t)/\delta (0)$). 
But at finite v, we have 
\be
\frac{\partial X /\partial z}{\partial X /\partial \tilde{r}}=
\gamma \frac{\gamma (z-vt)}{\tilde{r}}
\rightarrow \gamma \rightarrow \infty \;\;\; ({\rm if \;\; as\;\; }
v\rightarrow 1, \gamma (z-vt)\sim \tilde{r})
\ee

Let's check the Lorentz invariance of the field distribution by looking at 
its energy, and  proving $E=\gamma E_0$. From 
\be
S= - \int d^3 x \;dt (\sqrt{1+(\partial_{\mu} X)^2}-1)
\ee
we find 
\be
H=\int d^3 x[\frac{1+(\vec{\nabla}X) ^2}{\sqrt{1+(\vec{\nabla}X)
 ^2-\dot{X}^2}}-1]
\ee
Then before the boost, 
\be
E_0= \int 4\pi r^2 dr [\sqrt{1+\frac{\bar{C}^2}{r^4-\beta^2 \bar{C}^2}}-1]=
\int 2\pi \tilde{r}d\tilde{r}dz[\sqrt{1+\frac{\bar{C}^2}{r^4-\beta^2 \bar{C}
^2}}-1]
\ee
and after the boost (changing the variable of integration from $z$ to 
$z'=\gamma (z-vt)$), we get 
\be
E= \frac{1}{\gamma} \int 2\pi \tilde{r}d\tilde{r}dz'
[\sqrt{1+\frac{\bar{C}^2}{r^4-\beta^2 \bar{C}^2}}\frac{1+ \frac{\bar{C}^2}
{r^4-\beta^2 \bar{C}^2}
\frac{\tilde{r}^2+\gamma^2 z'^2}{r^2}}{1+\frac{\bar{C}^2}{r^4-\beta^2 \bar{C}
^2}}-1]
\ee
where $r^2=\tilde{r}^2+z'^2$. We can see at most that at large $\gamma$, 
E goes like $\gamma^2/\gamma =\gamma$, as it should.

{\bf Boosted BIon}

Now the scalar is actually the electric field, the zeroth component of a 
vector, thus the action is 
\be
S= -\int d^3 x \; dt [\sqrt{1-(\vec{\nabla} \phi- \dot{\vec{A}})^2
+(\vec{\nabla}\times \vec{A})^2- ((\vec{\nabla} \phi- \dot{\vec{A}})\cdot
(\vec{\nabla}\times \vec{A}))^2}-1]
\ee
and becomes
\be
S= -\int d^3 x \; dt [\sqrt{1-(\partial \phi)^2}-1]
\ee
 only when $\vec{A}=0, \dot{\phi}=0$, thus we have to boost
differently, taking into account that $\phi $ is not a scalar anymore, but 
now $(\phi, \vec{A})$ is a vector. 

Then on a static, purely electric configuration like the BIon
($F_{0i}=E_i, F_{ij}=\epsilon_{ijk}B_k$) we get 
\bea
&&E'_z(x', y', z', t')= E_z(x, y, z);\;\; 
E'_x(x_{\mu}')=\gamma E_x(x_{\mu});\;\; E'_y(x_{\mu}')= \gamma E_y(x_{\mu})
\nonumber\\&& B'_y(x_{\mu}')= \gamma v E_x(x_{\mu});
\;\; B'_x(x_{\mu}')= -\gamma v E_y(x_{\mu});\;\; B'_z=0\label{transform}
\eea
Taking  $E_i=-\partial_i f(\sqrt{x^2
+y^2+z^2})$ and boosting it to obtain $E'_{i'}=
\partial _{t'} A'_{i'}-\partial_{i'}g(x', y, z', t')$, and if we also 
boost $(f, \vec{A})$ as a vector (i.e. giving 
an extra condition, that we don't change gauge when boosting, which is 
a nontrivial condition, as this is not the case for gravity, see below), 
we obtain
\be
g(x, y, z, t)= A_0'=\gamma(A_0+v A_z)= \gamma A_0 = \gamma 
f(\sqrt{x^2+y^2+\frac{(z-vt)^2}{1-v^2}})
\ee

Specifically, for the BIon we have 
\be
E'_z=\frac{C}{\sqrt{r^4+\beta^2 C^2}}\frac{\gamma (z-vt)}{r}; \;\;\;
 E'_{\tilde{r}}
=\frac{C}{\sqrt{r^4+\beta^2 C^2}}\frac{\gamma \tilde{r}}{r}; \;\;\;
 B'_y=\frac{C}{\sqrt{r^4+\beta^2 C^2}}\frac{\gamma v x }{r}
\ee
thus in the large $\gamma $ limit,
\be
\frac{E'_{\tilde{r}}}{E'_z}=\frac{\tilde{r}}{\gamma (z-vt)}\cdot 
\gamma \rightarrow \infty \;{\rm if}\; \gamma (z-vt)\sim \tilde{r};\;\;\;
B'_{\theta}\rightarrow E'_r
\ee
If we boost $\phi$ as the 0 component of a vector, at $v\rightarrow 1$
we get that $\phi'=0$ if $z\neq t$, but now $\phi(z=t)\sim \gamma \rightarrow 
\infty$. Moreover, the width in $z-vt$ is $\sim 1/\gamma$, thus 
\be
\phi '(\tilde{r}, z, t)=a\phi (\tilde{r})\delta (z-t)
\ee
with $a$ a number.   

The extremal BIon of Callan and Maldacena \cite{cm} is a  solution to 
the action in (\ref{daction}).
The static BPS solution is 
\be
X=\frac{q}{|\vec{r}-\vec{r}_0|^{p-2}};\;\;\; \vec{E}=\vec{\nabla} X
\ee
thus both X and $\vec{E}$ are singular at r=0. Boosting this will generate 
the same $X$ as for the catenoid (with horizon at r=0),
 but now we will have both electric and 
magnetic fields as well (thus the action in (\ref{daction})
is not valid anymore, it 
needs to be ``boosted'' as well, i.e. the $\vec{B}$ dependence specified).

Note that when we boost the black hole solution to obtain the A-S solution, 
we write 
\be
g_{tt}(x', y', z', t'))= 1-\frac{2MG}{\sqrt{x'^2+y'^2+\frac{(z'-vt')^2}{
1-v^2}}};\;\; g_{rr}=1/g_{tt}
\ee
and then we transform $g_{tt}, g_{rr}$ to $g'_{\mu'\nu'}$ (and take the 
limit $v\rightarrow 1$), which would be the equivalent of calculating only 
$\vec{E}'$, not $g$. Moreover, one then has to transform the coordinates 
also to reach the system where the A-S looks simple. It could be that the 
same is required here (for BIons and catenoids), 
but we will not pursue this further.

\section{BIon solutions and ``dumb holes'' as fireballs}

In \cite{unruh} it was found that there are analogs of black holes in 
hydrodynamics, dubbed ``dumb holes''. When a configuration of fluid moves 
at ultrasonic speed it creates horizons that radiate thermally, analogous 
to the thermal horizons of black holes. The hydrodynamics equations are 
written for a potential flow, in terms of a scalar potential $\Phi$. 
On the other hand, we want to obtain also a thermal horizons in the collision 
of scalar ($\Phi$) shockwaves in DBI theory. We will see that in fact the 
equations in the two cases are completely similar. We will thus first review 
the ``dumb holes'' and then completely parallel the calculation for the 
DBI scalar, calculating the temperature of scalar ``fireballs''. 
For the ``fireballs'' we will take the static spherically symmetric solutions 
studied in section 3, and we will discuss towards the end whether this is a 
good approximation to the real dynamical situation.  

For ultrasonic fluid flow, the surface where $v=c$ 
(the velocity of particles reaches the
sound velocity) is a horizon that radiates particles thermally, and can be 
mapped to a black hole. The fluid has an equation of state $p= p(\rho)$ and
the fluid motion is irrotational $\vec{\nabla}\times \vec{v}=0$, such that 
one can write $\vec{v}=\vec{\nabla}\Phi$ (potential flow). The speed of sound 
is defined by $c^2=dp/d\rho$. 

The equations of motion are the local pressure equation and the continuity 
equation, i.e.
\be
\rho (\dot{\vec{v}}+(\vec{v}\cdot \vec{\nabla})\vec{v})= -\vec{\nabla} p 
(\rho);
\;\; \dot{\rho}+\vec{\nabla}\cdot (\rho \vec{v})=0
\ee
The first equation is integrated to the Bernoulli equation
\be
\dot{\Phi}+\frac{\vec{v}\cdot \vec{v}}{2}+h(\rho)=0;\;\;\; h(\rho)=\int 
\frac{dp}{\rho}
\ee
If we now calculate the fluctuation equations, in variables $\phi=\delta 
\Phi$ and $\psi= \delta \rho/\rho$, we get 
\bea
&&\dot{\phi}+\vec{v}\cdot \vec{\nabla} \phi + c^2 \psi=0\nonumber\\
&& (\frac{d}{d t} +\vec{v}\cdot \vec{\nabla}) \psi +(\vec{\nabla} \phi)\cdot 
\vec{\nabla}\;  ln \; \rho+\vec{\nabla}^2\phi =0
\eea
Eliminating $\psi$ we get 
\be
\frac{1}{\rho}(\frac{d}{dt}+\vec{v}\cdot\vec{\nabla}+(\vec{\nabla}\cdot\vec{v}
))\frac{\rho}{c^2}(\frac{d}{dt}+\vec{v}\cdot\vec{\nabla})\phi-\frac{1}{\rho}
\vec{\nabla}( \rho \vec{\nabla}\phi)=0
\ee
which is exactly the equation of motion for a scalar field in a 
curved spacetime, $\partial_{\mu}
\sqrt{g}g^{\mu\nu}\partial_{\nu} \phi=0$ if the metric is given by 
\be
\sqrt{g}g^{\mu\nu}=\rho\begin{pmatrix} \frac{1}{c^{2}} & \frac{v^i}{c^2}\\
\frac{v^j}{c^2} & \frac{v^i v^j}{c^2}-\delta ^{ij}\end{pmatrix}
\label{metricmap}
\ee
which implies in 4d (after finding $g_{\mu\nu}$ and defining a new time 
coordinate by $d\tau=dt +v^i dx^i/(c^2-v^2)$)
\be
ds^2= \frac{\rho}{c}
[ (c^2-v^2)d\tau^2-(\delta^{ij}+\frac{v^i v^j}{c^2-v^2})dx^i dx^j]
\ee
or, in the case of a radial flow
\be
ds^2=\frac{\rho}{c}
 [ (1-v^2/c^2)c^2d\tau^2-\frac{dr^2}{1-v^2/c^2 }-r^2 d\Omega^2]
\label{radflow}
\ee
In the new coordinates, the scalar wave equation is 
\be
\partial_0 \frac{\rho/c^2}{1-v^2/c^2}\partial_0\phi +\partial_i
\rho (\frac{v^i v^j}{c^2}-\delta^{ij})\partial_j \phi =0
\label{neweq}
\ee

At constant (or neglijible variation of) $\rho$, the fluctuation equation 
in the original coordinates can be written as 
\be
[\partial_t +(\vec{\nabla}\Phi)\cdot \vec{\nabla}+\vec{\nabla}^2 \Phi]
[\partial_t+(\vec{\nabla}\Phi)\cdot \vec{\nabla}]\delta \Phi - c^2 \vec{
\nabla}^2 \delta \Phi=0
\ee

The horizon is the surface where $v=c$ ($(\nabla\Phi)^2 =c^2$), 
thus the metric becomes singular and then in the above equation we have 
\be
[\partial_t^2 + ((\vec{\nabla}\Phi)^2-c^2) \vec{\nabla}^2 +...]\delta \Phi=0
\ee
thus the coefficient of $\vec{k}^2 $ in a fluctuation mode $\vec{k}$ 
changes sign. 

For static fluctuations ($\partial_t=0$), the equation is 
\be
-c^2 \nabla^2 \delta \Phi+(\nabla^2 \Phi) (\vec{\nabla}\Phi)\cdot \vec{\nabla}
\delta \Phi + (\vec{\nabla}\Phi)\cdot \vec{\nabla}((\vec{\nabla}\Phi)\cdot
\vec{\nabla} \delta \Phi)=0
\ee

{\bf DBI Lagrangean}

The DBI scalar has a Lagrangean
$\sqrt{1+(\vec{\nabla}\Phi)^2-(\partial_t \Phi)^2}$. Here we 
denote the scalar by $\Phi$ just to underline the analogy with the 
hydrodynamics case. It gives 
the fluctuation equation on static solutions ($\partial_t \Phi=0, 
\partial_t\delta \Phi \neq 0$):
\bea
&&\delta [\vec{\nabla}\frac{\vec{\nabla}\Phi}{\sqrt{1+(\vec{\nabla}\Phi)^2}}]
-\frac{1}{\sqrt{1+(\vec{\nabla}\Phi)^2}}\partial_t^2 \delta \Phi\nonumber\\
&&\simeq -\frac{1}{\sqrt{1+(\vec{\nabla}\Phi)^2}}\partial_t^2 \delta \Phi+
\vec{\nabla}\cdot (\frac{\vec{\nabla}\delta \Phi}
{\sqrt{1+(\vec{\nabla}\Phi)^2}})-  \nonumber\\&&-
\vec{\nabla}\cdot [ \frac{\vec{\nabla}\Phi}
{(1+(\vec{\nabla}\Phi)^2)^{3/2}}(
(\vec{\nabla}\Phi)\cdot \vec{\nabla}\delta \Phi)]=0
\label{fluctu}
\eea

We see that now the equation is very similar to the one above. The horizon 
is again where the coefficient of $\nabla^2 \delta \Phi$ changes sign, 
which now means 
\be
\frac{1}{\sqrt{1+(\vec{\nabla}\Phi)^2}}= \frac{(\vec{\nabla}\Phi)^2}{
(1+(\vec{\nabla}\Phi)^2)^{3/2}}
\ee
whose solution is $|\nabla \Phi|=\infty$, which is what we called a horizon 
for the ``catenoid'' anyway! Before we thought of the  $|\nabla \Phi|=\infty$
solution as a horizon just because of the singularity, but now we see that 
it is the exact analog of the dumb hole horizon, thus of the black hole 
horizon. 

In fact, we can do more. The fluctuation equation (\ref{fluctu}) can be 
rewritten in a suggestive form as (using that $\partial_t \Phi=0$)
\be
-\partial_t \frac{1}{\sqrt{1+(\vec{\nabla}\Phi)^2}}\partial_t \delta \Phi
+ \partial_i \frac{1}{\sqrt{1+(\vec{\nabla}\Phi)^2}}(\delta^{ij}-
\frac{\partial^i \Phi}{\sqrt{1+(\vec{\nabla}\Phi)^2}}
\frac{\partial^j \Phi}{\sqrt{1+(\vec{\nabla}\Phi)^2}})\partial_j \delta \Phi=0
\ee
which is the same as the black hole equation in the new coordinates 
(\ref{neweq}) with the identification 
\be
c^2=1+(\vec{\nabla}\Phi)^2, \;\;\; \rho=\frac{1}
{\sqrt{1+(\vec{\nabla}\Phi)^2}};\;\;\; \frac{v^i}{c}=\frac{\partial^i \Phi}{
\sqrt{1+(\vec{\nabla}\Phi)^2}}\;\;({\rm i.e.,} \; v^i=\partial^i \Phi)
\label{velo}
\ee

Of course, just because v=c when $|\nabla \Phi |=\infty$ it is not a 
good reason for it to be a thermal horizon. There are many hydrodynamic 
surfaces where $v=c$ that do not radiate thermally as ``dumb holes'', 
just as there are horizons of finite area, for extremal black holes, that 
nevertheless are at T=0. The key point is that the variation at the 
horizon, $dv/dr|_{v=c}$ for dumb holes and $\partial_r g_{tt}|_{r=r_H}
\propto k$ for black holes that gives the temperature of the horizon. But the 
point is that neither for the black holes or the dumb holes, the presence 
of a singularity is not required, the thermality is a property of the 
horizon itself. What lies behind the horizon is of no consequence. 

{\bf Information trapping and horizons}

One should observe though that the horizon of the scalar fireball doesn't 
have the black hole horizon property that nothing can escape from the 
horizon. It has that property only for scalar field excitations
and only in a very limited sense.  In general, excitations 
cannot come out of the horizon because of the same reason that that 
the horizon  radiates 
thermally (the coefficient of $\vec{\nabla}^2 
\phi$, proportional to $v^2-c^2$, 
becomes zero at the horizon in a specific way). This is seen as follows.
 
For a Schwarzschild black hole, the infinite time delay necessary for a 
massless particle to come out, from the perspective of an outside observer 
is due to equating $ds^2=0$ and finding that then 
\be
\int dt = \int^R \frac{dr}{c(1-2M/r)}\rightarrow \infty{\rm \;\; as\;\;}
R\rightarrow r_H
\ee
For the metric in (\ref{radflow}) this becomes $\int dr/[c(1-v^2/c^2)]$ being 
divergent at the horizon, which is easily seen to be due to the fact that 
$d[c(1-v^2/c^2)]/dr$ is a nonzero constant at the horizon. This is
 the same condition 
that we will find below for the nonzero horizon temperature, if $\rho$ 
is a constant or satisfies $\rho c (1-v^2/c^2)\neq $ constant
 (in our case $\rho$ is not constant but satisfies the latter property). 
 Thus the infinite time delay and finite
nonzero temperature are almost always related, as expected, and if it's true,
 scalar excitations cannot get out. However, in our case, $d[c(1-v^2/c^2)]/dr$
is infinite at the horizon, implying a finite time delay, and we will find 
below an infinite temperature as well. 

But any other (nonscalar) excitation can always come out. We 
might ask how come $\Phi$ fluctuations can't get out
if $d[c(1-v^2/c^2)]/dr$ is nonzero and finite at the horizon, 
after all, we are in 
flat space so it takes a finite time for radiation to come from the 
horizon. But the point is that the scalar fluctuation equation is the same as
for a black hole, meaning the {\em characteristic} equation for a scalar 
field perturbation will propagate as a function of the asymptotic 
coordinates $r,t$ the same way as from a black hole. In other words, the 
phase and group velocities ($c_{ph}$ and $c_{gr}$) of scalar excitations
 would tend to zero in those cases and information could not be exchanged 
with the inside of the horizon.

In our case, substituting $\delta \Phi= A \; exp(i(\omega t -kr))$ 
(spherical waves) in the perturbation equation (\ref{fluctu}) one obtains
\be
\omega^2 = \frac{1}{1+\Phi'^2}(k^2-3k \frac{\Phi' \Phi ''}{1+\Phi '^2})
= k^2 \frac{r^4-r_0^4}{r^4}+6k \frac{r_0^4}{r^5}
\ee
thus we see that the coefficient of $k^2$ becomes zero at the horizon and 
the coefficient of $k$ is finite. One can easily check that if for our solution
we had $\Phi ' \sim 1/(r-r_0)$ near the horizon (which would imply both finite 
nonzero temperature and  infinite ``geodesic'' 
time delay) we would get also the 
coefficient of $k$ to go to zero at the horizon, but slower than the 
coefficient of $k^2$. As it is, we get the phase and group velocities
\bea
&&c^2_{ph}=\frac{\omega^2}{k^2}=\frac{r^4-r_0^4}{r^4}+\frac{6}{k}\frac{r_0^4}{
r^5}\equiv a+\frac{b}{k}
\rightarrow \frac{6}{k r_0} \; \; {\rm as\; r\rightarrow r_0}
\nonumber\\&&
c_{gr}= \frac{d\omega }{dk}= \frac{a+b/(2k)}{a+b/k}\rightarrow \frac{1}{2}
\sqrt{\frac{b}{k}}\rightarrow \frac{1}{2} \sqrt{\frac{6}{kr_0}}
 \; \; {\rm as\; r\rightarrow r_0}
\eea
and we see that infinite k modes have zero phase and group velocities. 
This is not a relativistic formula ($\omega =c k$, c= constant), but not 
a nonrelativistic one either (which would generically be of the type 
$c=c_0 +a k+ bk^2+...$), but comes from the gravitational (black hole) 
background.
The fact that one could apparently have $c_{ph}>1$ and $c_{gr}>1$ is an 
artefact of our approximation of spherical waves, which clearly only 
works at the horizon if $k\gg 1/r_0$ ($\lambda\ll r_0$). 

Thus high energy perturbations have phase and group velocities going to zero 
as $1/\sqrt{k}$, hence for them the horizon is indeed impenetrable, and 
information can be exchanged with increasing difficulty with the inside 
of the horizon. If we would have infinite ``geodesic'' time delay 
($\Phi ' \sim 1/(r-r_0)$ near the horizon), then the phase and group velocities
would be zero at the horizon, thus all scalar information would be stuck 
at the horizon, exactly as for a black hole. Then the phase and group 
velocities would be again proportional to $1/\sqrt{k}$, but would be multiplied
by a factor that vanishes at the horizon. 

Finally, notice that if we take the original electromagnetic DBI action 
for $\phi$ (for the BIon, with the 
minus sign in the square root), the sign of the two terms in (\ref{fluctu})
is the same, thus one can never cancel them against each other. Thus 
the original DBI action never admits horizons!

{\bf Temperature calculation}

For a black hole in flat space, the horizon temperature is 
\be
T=\frac{k}{2\pi}
\ee
where $k$ is the ``surface gravity'' of the horizon. For a static, spherically
symmetric solution with only $g_{rr}(r)$ and $g_{tt}(r)$ nontrivial
(and possibly an r-dependent conformal factor for the sphere metric), one can 
easily calculate that 
\be
(2k)^2= \lim_{horizon} \frac{g^{rr}}{g_{tt}}(\partial_r g_{tt})^2
\ee
For a Schwarzschild black hole, $g^{rr}=g_{tt}$ and 
we calculate that $k=1/(4MG)$, as known. For the ``dumb hole'', using 
the above map to a curved spacetime (\ref{metricmap}) we get that 
\cite{unruh}
\be
(2k)^2 = \{\frac{1}{\rho}\partial_r[\rho c (1-v^2/c^2)]\}^2 |_{v=c}
\Rightarrow T=\frac{1}{4\pi}\frac{1}{\rho}\partial_r[\rho c (1-v^2/c^2)] 
|_{v=c}
\ee
Note that in Unruh's case, where $\rho$ and $c$ are nonzero and finite
at the horizon, 
one gets $T=(dv/dr|_{v=c})/(2\pi)$, but in our case that is not true.

For the catenoid, we have 
\be
\Phi'= \frac{D}{\sqrt{1-D^2}}= \frac{\bar{C}/r^2}{\sqrt{1-(\bar{C}/r^2)^2}}
\ee
and as we saw we map to the ``dumb hole'' calculation by defining the 
velocities (\ref{velo})
\be
v^2/c^2= [\frac{(\vec{\nabla}\Phi)^2}{(1+(\vec{\nabla}\Phi)^2)}]
;\;\;\; c^2= 1+\Phi '^2 ;\; \rho = 1/\sqrt{1+\Phi '^2}
\ee
and then 
\bea
&&2k=|\sqrt{1+\Phi '^2}\frac{d}{dr}[\frac{1}{1+\Phi'^2}]|_{r=r_0}=
|\sqrt{1+\Phi '^2}\frac{d}{dr} [\frac{
\bar{C}^2}{r^4}]|_{r=r_0}= \sqrt{\frac{r_0}{4(r-r_0)}}
\frac{4}{r_0}\nonumber\\
&&\Rightarrow T= \sqrt{\frac{r_0}{4(r-r_0)}}
\frac{1}{\pi r_0}
\label{temper}
\eea
thus the temperature is infinite!

Obviously, the temperature of a system cannot be infinite. There are several 
possible interpretations. First, we notice that the infinite square root 
factor in front of the temperature comes from $\sqrt{1+\Phi '^2}$, which 
is equal to c using the map to black holes, and if c was a constant, we 
could have rescaled the time coordinate to get rid of it in the wave 
equation. As it is, we cannot rescale it away. 
But $\sqrt{1+\Phi'^2}$ is also the energy density 
of the solution, which cannot become infinite, so there will either be 
higher order corrections to the action preventing that, or else for the real 
dynamical (time dependent) process of high energy collisions one cannot  
approximate by the production of a  static spherically symmetric solution.
Thus at the classical level, the temperature could be made finite either 
by higher order corrections in the action, or by deviations from time 
independence and sphericity. As an example, the fluctuation equation around a
time dependent solution is 
\bea
&&\partial_i \frac{1}{\sqrt{1+(\vec{\nabla}\Phi)^2-\dot{\Phi}^2}}(\delta^{ij}-
\frac{\partial^i \Phi}{\sqrt{1+(\vec{\nabla}\Phi)^2-\dot{\Phi}^2}}
\frac{\partial^j \Phi}{\sqrt{1+(\vec{\nabla}\Phi)^2-\dot{\Phi}^2}})
\partial_j \delta \Phi\nonumber\\
&&
+\nabla^i  \frac{1}{\sqrt{1+(\vec{\nabla}\Phi)^2-\dot{\Phi}^2}}
\frac{\nabla ^i \Phi \dot{\Phi}}{1+(\vec{\nabla}\Phi)^2-\dot{\Phi}^2}
\partial_t \delta \Phi \nonumber\\&&
+\partial_t  \frac{1}{\sqrt{1+(\vec{\nabla}\Phi)^2-\dot{\Phi}^2}}
\frac{\nabla ^i \Phi \dot{\Phi}}{1+(\vec{\nabla}\Phi)^2-\dot{\Phi}^2}
\nabla^i \delta \Phi\nonumber\\
&&-\partial_t \frac{1}{\sqrt{1+(\vec{\nabla}\Phi)^2-\dot{\Phi}^2}}(1+
\frac{\dot{\Phi}^2}{1+(\vec{\nabla}\Phi)^2-\dot{\Phi}^2})\partial_t \delta
\Phi =0
\eea
which will have a modified map to the black hole fluctuation equation. 
Finally, besides all this possible classical resolutions, an infinite 
temperature implies an infinite particle production, thus quantum mechanics
could also regulate this behaviour. In any case, we will assume that the 
infinite factor $\sqrt{1+\Phi '^2}$ which is proportional, as noted, to the 
energy density of the solution, is regulated to some large but finite value, 
and calculate the mass scaling of the temperature coming from $r_0$.

Restoring the energy scale $\hat{M}_P$, we get 
\be
T\propto \frac{1}{ r_0}=
\frac{1}{ \sqrt{\bar{C}}}\propto \frac{\hat{M}_P^{4/3}}{M^{1/3}}
\ee
($\sqrt{\bar{C}}$ is related to the mass of the catenoid by $M=I \bar{C}^{
3/2}/g$, as we saw in section 3.) 

This calculation is valid as long as $r_0<
m_{\phi}$, thus for $M<\hat{M}_P^4/m_{\phi}^3$. After that we have to take 
$m_{\phi}$ into account. Similarly we then get 
\be
T=\frac{1}{4\pi} |\sqrt{1+\Phi'^2+m^2\Phi^2}
\frac{d}{dr}\frac{1+m^2{\Phi}^2}{1+{\Phi '}^2 +m^2 \Phi^2}|_{
r=r_1}
\ee
Near the horizon, the equation of motion for the massive DBI scalar becomes 
\be
r\Phi '' (1+ m^2 \Phi ^2)+2\Phi '^3=0
\ee
giving the approximate solution
\be
\Phi (r\simeq r_1)\simeq \Phi_1 +\sqrt{r_1(1+m^2 \Phi_1^2)(r-r_1)}
\ee
This gives then 
\be
T \simeq \sqrt{1+m^2 \Phi_1^2}\sqrt{\frac{r_1}{4(r-r_1)}}\frac{1}{\pi r_1}
\ee

At large distances, 
$\Phi \sim \bar{C}e^{-mr}/r$, with $\bar{C}$ being some power law of the mass. 
Thus perturbatively, the horizon will be where $\Phi$ becomes of order 1, 
and if we would apply this perturbative formula we would get 
$r_1\sim 1/m \ln(\bar{C}m)$.  Considering again that the infinite factor
$\sqrt{1+\Phi'^2 +m^2}= \sqrt{1+m^2\Phi_1^2}\sqrt{r_1/(4(r-r_1))}$,
proportional to the energy density at the horizon, will be regulated to 
a finite value and calculating only the mass scaling with $r_1$  we obtain 
\be
T\propto \frac{m}{\ln(\bar{C}m)}\propto m
\ee
as we argued in \cite{nastase3} we should get. 

Finally, if the temperature does become finite as we argued, by the  
arguments we already gave, light will take an infinite time to reach 
the horizon, a fact observed from geodesic arguments in the black hole 
background and from the vanishing of $c_{ph}$ and $c_{gr}$ in the scalar 
background.

\section{SU(2) DBI actions and Pavlovskii's topological soliton}

Based purely on phenomenological grounds, Pavlovskii \cite{pav} 
looked at a DBI action for the SU(2) pions of QCD. It contains an 
infinite number of higher derivative corrections to the nonlinear sigma 
model action, that have a chance of having a Skyrme-like topological soliton. 
Such a soliton was indeed found. Here we will extend that discussion and 
look for modifications of the SU(2) actions that can fit our purposes.
We will carefully look at the numerical solutions in order to generalize 
in the next section to the case of interest.

The action \cite{pav} writes is 
\be
{\cal L}= f_{\pi}^2 Tr \beta^2 [\sqrt{1+\frac{1}{2\beta^2}L_{\mu}L^{\mu}}
-1]
\label{pavlo}
\ee
where $L_{\mu}= U^{-1}\partial_{\mu} U$, $U= exp(i\vec{\phi}_{\pi}\vec{\tau}
/f_{\pi})$, and chooses a spherically symmetric configuration (``hedgehog'')
\be
U= e^{i F(r)(\vec{n}\vec{\tau})};\;\;\; \vec{n}=\frac{\vec{r}}{r}
\ee
One should note that the sign inside of the square root is different in 
\cite{pav} due to a local sign mistake. In the spherically symmetric 
solution used afterwards he has the same signs in terms of F, 
which is correct but 
because of the i's in U, the sign changes between the 
above action and the action in terms of F (i.e. $L_iL_i=-F'^2+...$). 

Then 
\be
U= e^{i \vec{n}\cdot \vec{\sigma} F(r)}= \cos F(r)+ i \vec{n}\cdot 
\vec{\sigma} \sin F(r)
\ee
and then for such a static configuration 
\bea
&& L_i= U^+\partial_i U= i n_i \vec{n}\cdot \vec{\sigma} (F'-\frac{\sin F 
\cos F}{r})+ i \sigma_i \frac{\sin F \cos F}{r}- i \frac{\sin ^2 F}{r} 
n_j \epsilon _{ijk }\sigma_k\Rightarrow \nonumber\\
&& L_i^2 = -F'^2 -2\frac{\sin^2 F}{r^2}
\eea
and the energy functional is ($E=-S$ for static solutions)
\be
E= 8 \pi f_{\pi}^2 \beta^2 \int _0^{\infty} (1-R)r^2 dr ;\;\;\;
R= \sqrt{1-\frac{1}{\beta^2}(\frac{F'^2}{2}+\frac{\sin^2 F}{r^2})}
\ee
And the equation of motion is 
\be
(r^2 \frac{F'}{R})'=\frac{\sin 2F}{R}
\ee

Explicitly, it gives
\bea
&&(r^2-\frac{1}{\beta^2 }\sin ^2 F )F''+ (2r F' -\sin 2F) \nonumber\\
&& -\frac{1}{\beta^2} (r F'^3-F'^2 \sin 2F + \frac{3}{r} F' \sin ^2 F
-\frac{1}{r^2} \sin 2F \sin ^2 F)=0\label{eom}
\eea

For a good solution, F should go to zero at infinite r, thus the 
large r expansion of such a solution is found to be  
\be
F(r)= \frac{a }{r^n}+\frac{b}{r^{n+m}}+\frac{c}{r^{n+p}}
\ee
and then from the equations of motion
\be
F(r)=  \frac{a}{r^2}-\frac{a^3}{21 r^6}-\frac{a^3}{3\beta^2 r^8}
\ee

The first 2 terms in F(r) appear just from the equation $r^2 F''
+2rF' - \sin 2F=0$, i.e. would appear also from the ``linear'' action 
(from the SU(2) sigma model)
\be
S= \int [\frac{(\partial F)^2}{2}+\frac{\sin ^2 F}{r^2}]=
 -\int Tr\frac{L_i^2}{2}
\ee
The first term appears when we approximate the sin to first order, and the 
second when we approximate to second order.

By comparison, for the single scalar DBI action, the first two terms in 
the expansion would come from the equation 
$r^2 F''+2r F' -r F'^3/\beta^2$, with the solution
\be
F=\frac{a}{r}-\frac{a^3}{20 \beta^2 r^5}
\ee

 Notice that in first order, 
\be
L_{\mu}L_{\mu}\simeq -(\vec{\sigma}\cdot \partial_{\mu}\vec{\pi})^2 = 
\partial_{\mu}\pi_i\partial_{\mu}\pi_i
\ee
which gives on a spherically symmetric solution (the above ansatz)
\be
-F'^2 -2 \frac{F^2}{r^2}
\ee
Thus for three perturbative scalars, with action 
\be
S=\frac{1}{2}\int d^4 x \partial_{\mu}\pi^i \partial_{\mu} \pi^i
\ee
if one chooses a spherically symmetric 
(``hedgehog'') solution, with $\pi_i=n_i F(r), 
\vec{n}=\vec{r}/r$, one gets
\be
S=\frac{1}{2}\int d^4 x (F'^2 +2 \frac{F^2}{r^2})
\ee
thus the difference in behaviour at infinity between the SU(2) case and the 
single scalar case comes not from the nonlinearity of the action (it 
couldn't), but rather from the spherically symmetric ansatz. 

The above action has solutions that have $F(0)=N\pi$. Such a solution
with $N\geq 1$, if 
it also goes to zero at infinity, will have a nonzero topological charge 
\be
B=\frac{1}{24\pi^2}\int d^3x \epsilon_{ijk}Tr(L_iL_jL_k)
\ee

If $F(0)=N\pi$, then we can find from the equations of motion that 
$F'(0)=a$ is not constrained, thus parametrizing a set of solutions
with $F=N\pi +ar+o(r^3)$. 
However, for a single value of $a$ (negative)
is the solution going to 0 at infinity, thus being topological. 

In \cite{pav}, the topological solution was found numerically, by 
imposing the good behaviour at infinity. For most solutions, one encounters 
a point where the solution becomes numerically unstable, at a nonzero $r=r_0$
satisfying $F(r_0)= \pm arcsin (\beta r_0)$, where we can check that 
$F'(r_0)=0$. Physically, it is clear that thes solutions will be continued to 
$F(0)=0$, $F'(0)>0$, thus carrying no topological charge. 

\begin{figure}

\begin{center}

\includegraphics{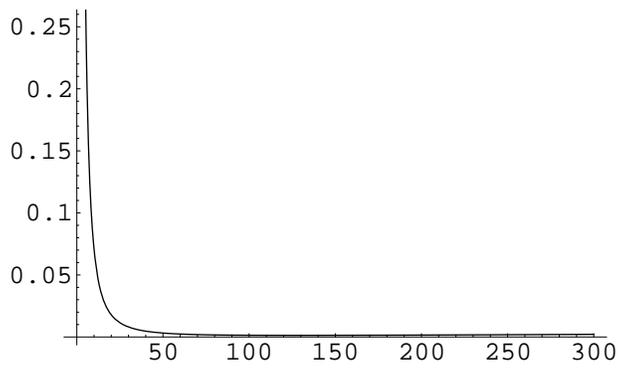}
\end{center}
\caption{Topological solution}
\label{sol3}
\end{figure}

\begin{figure}

\begin{center}

\includegraphics{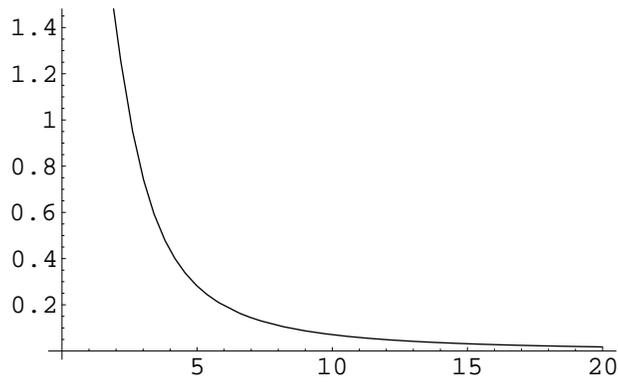}
\end{center}
\caption{Topological solution, detail}
\label{sol2}
\end{figure}

We have also analyzed numerically what happens when the asymptotics at 
zero are correct, i.e. $F(0)=\pi$ (the other N's are similar). 
Numerically, by trial and error, 
(everything is in units of $\beta$)
we find that the topological solution  has $a\simeq -0.809466$. 
Specifically, for $F(0.01)=\pi -0.00809466, F'(0.01)= -0.809466$. Then 
$F(300)\simeq 0$. Specifically, $F(10)= 0.0706502= b/r^2\Rightarrow 
b=7.065$ and then $F'(10)=-0.014102\simeq -2b/r^3$, but the relation is 
not true anymore at r=100, showing that this is an approximate solution.
This topological solution is given in fig.\ref{sol3}, with a detail in 
fig.\ref{sol2}.

Note that this value for $a $ is very  close to the value for which the 
square root argument becomes negative, thus the energy becomes imaginary:
\be
F\simeq \pi -a r+...\Rightarrow
\sqrt{1-(\frac{F'^2}{2}+\frac{\sin^2 F}{r^2})}\simeq \sqrt{1-\frac{3a^2}{2}}
\Rightarrow |a|\leq \sqrt{\frac{2}{3}}=0.8164965809... 
\ee

\begin{figure}

\begin{center}

\includegraphics{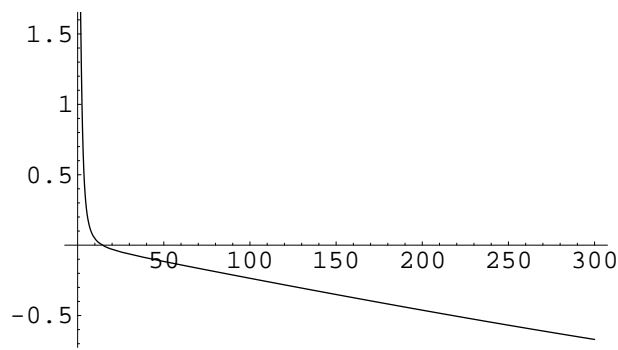}
\end{center}
\caption{Slightly below topological  solution}
\label{pavsign3}
\end{figure}

\begin{figure}

\begin{center}

\includegraphics{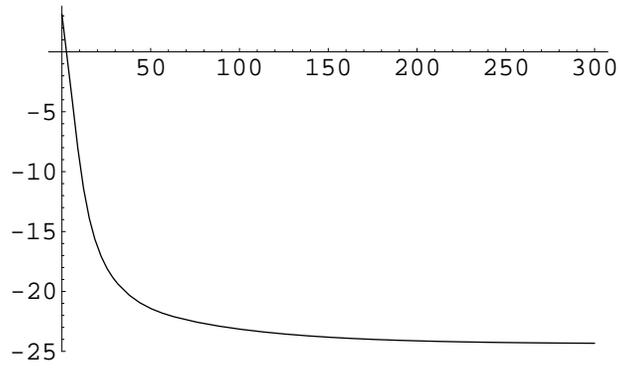}
\end{center}
\caption{Limit solution}
\label{pavsign4}
\end{figure}

For $a$ between the topological solution 
value $a\simeq -0.809466$ and the minimum 
value $a=-0.8164965809...$ F is negative at infinity, but the numerics are 
unstable, so it is not clear when it is actually reached. 
As an example, for $F(0.01)=\pi -0.008096, F'(0.01)= -0.8096$, 
$F(250)\simeq -0.5$ and falling (see fig.\ref{pavsign3}), 
and for  $a=- 0.8164965$ we get 
$F(300)$ flattening out in the neighbourhood of $-25$
(see fig.\ref{pavsign4}). However, from the 
equation of motion, we see that the allowed values at infinity are only 
$F(\infty)=k\pi /2$, with k an integer. Of course, for odd k, we don't 
have a solution with topological charge. 

\begin{figure}

\begin{center}

\includegraphics{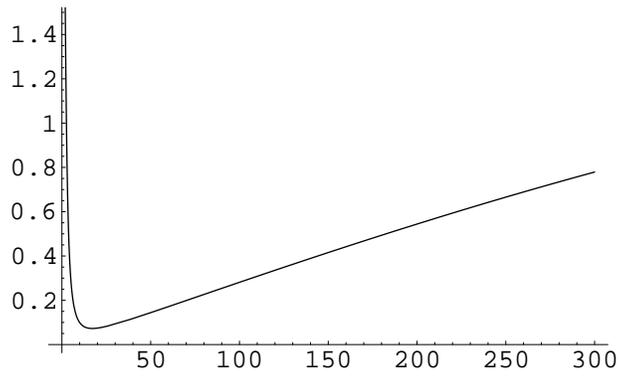}
\end{center}
\caption{Slightly above topological solution}
\label{pavsign2}
\end{figure}

For $a$ larger than the topological value, $F(\infty)$ is positive, and the 
same remarks apply. For example, for $a=-0.8093$, $F(300)\simeq 0.8$ and 
increasing (see fig.\ref{pavsign2}).

\begin{figure}

\begin{center}

\includegraphics{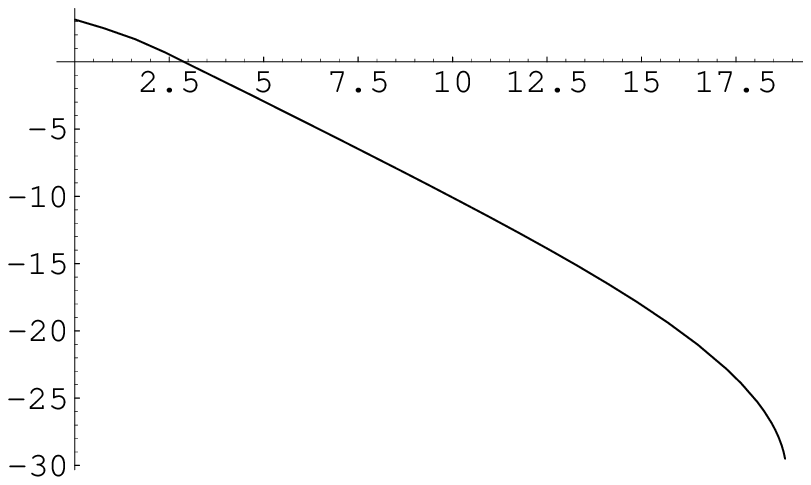}
\end{center}
\caption{Solution with imaginary energy}
\label{pavsign}
\end{figure}

Note that from the point of view of the equations of motion there is nothing 
wrong with $|a| > 0.8164965809... $, just that the energy (or action) of the 
solution is imaginary. Then numerically, we can check that we obtain 
horizons: F becomes more and more negative, until at a finite point 
$F'(r_0)=-\infty$. For example,  for $a=-0.905$ this happens at about 
$r=0.449$, for $a=-0.816502$ it happens for at about $r= 18.84$, etc.
(see fig.\ref{pavsign}). 

In conclusion, we see that Pavlovskii's action admits topological solutions.
Most solutions are not topological, thus unstable, and there are no solutions 
with horizons that have both real energy and good behaviour at infinity. 

\begin{figure}

\begin{center}

\includegraphics{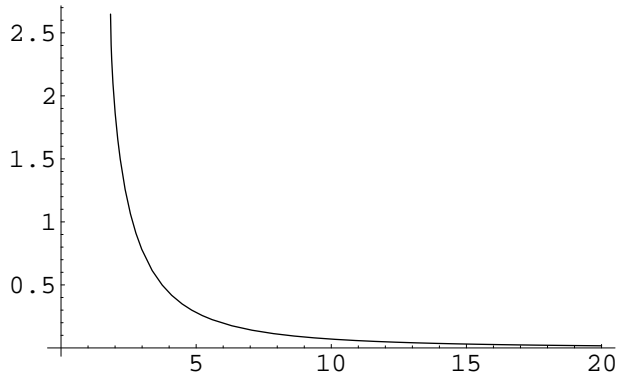}
\end{center}
\caption{Solution with horizon ($\beta^2<0$);
asymptotics at infinity of topological solution}
\label{oursign2}
\end{figure}

\begin{figure}

\begin{center}

\includegraphics{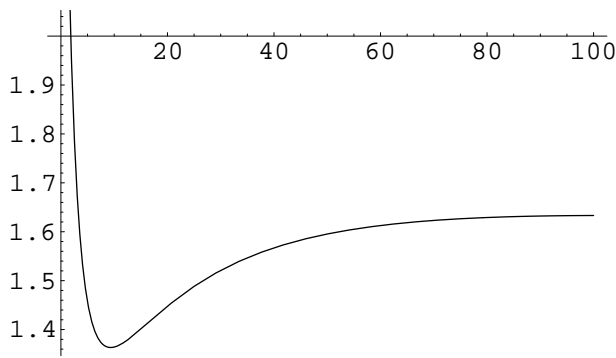}
\end{center}
\caption{The $\beta^2<0$ case;
asymptotics at zero of topological solution}
\label{oursign}
\end{figure}

One can ask what happens when we change the sign of $\beta^2$ in the 
action (\ref{pavlo}). We find 
that for the solution with the same asymptotics at infinity as Pavlovskii's 
topological solution (the first two terms in the asymptotics at infinity 
do not depend on $\beta^2$), i.e. with $b\simeq7.05$ at $r=10$, specifically 
$F(10)= 0.07, F'(10)=-0.0141$
(see fig.\ref{oursign2}), we get a horizon at $r=1.8310126459463$
(computer goes out of memory). Reversely, for the same asymptotics at zero 
as Pavlovskii's solution, i.e. with $a=-0.809466$, for our action, F 
goes down to about 1.36 and then goes back up to a maximum 
slightly above 1.6, asymptoting to $F(\infty)=
\pi/2=1.5707963268$ (see fig.\ref{oursign}).
The same qualitative behaviour (down to a minimum below $\pi /2$, 
up to a maximum above $\pi/2$
 and asymptoting to $\pi/2$) is valid for a whole range 
of values of $F'(0)=a$. We can easily check that $F'=0$ implies 
$F''(r_0)=sin(2F(r_0))/r_0^2$, thus we have minima below $F=\pi/2$ and 
maxima above it. 

\begin{figure}

\begin{center}

\includegraphics{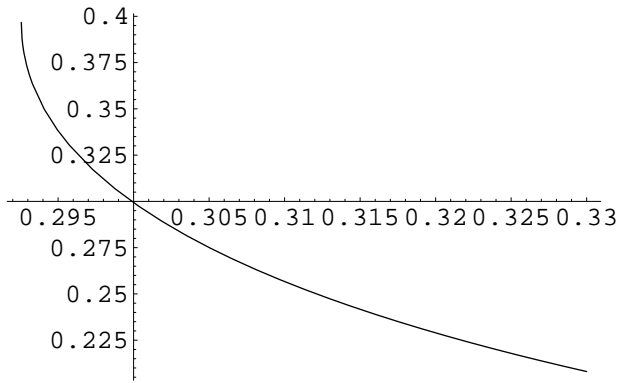}
\end{center}
\caption{Detail of horizon}
\label{oursign3}
\end{figure}

We have tried a large range of parameters $a$ (for asymptotics at r=0) and 
$b$ (for asymptotics at $r=\infty$), but always the same result: If the 
asymptotics at zero are good, then the solution tends to $\pi/2$ instead of 
0 at infinity, thus being non-topological, i.e. unstable. If the asymptotics 
at infinity are good, then the solution develops a horizon at finite r ($F'(r)$
is infinite, and within computer accuracy, F seems finite, thus
having a  square root singularity, as expected). 
A detail of the square root horizon, for $F(r\rightarrow \infty)\simeq 
0.002/r$ is given in fig.\ref{oursign3}.
We have tried values of b (at r=10) varying 
between 0.02 and 7.05. A few examples: b=0.02 has horizon at about
r=0.292, b=1 at about r=1.044, b=2 at about 1.290, b=4 at about 1.574. 

Let us now check under which conditions can we have a horizon, with F(r) finite
but $F'(r)=\infty$ at a finite $r=r_1$. From the equations of motion  
one finds 
\be
F(r\simeq r_1)\simeq F(r_1)+\sqrt{\frac{2(\sin^2 F(r_1)-\beta^2 r_1^2)}{r_1}
(r-r_1)}+...
\ee
We see that if $\beta^2>0$ (Pavlovskii's case), the solutions with good 
asymptotics at infinity indeed cannot have horizons. For them 
$(\sin^2 F(r_1)-\beta^2 r_1^2)<0$ at infinity and it becomes equal to 
zero at the point where the solution becomes indeterminate. 
Horizons can be obtained if $r<r_1$, i.e. $F'(r_1)=\infty$ from 
below, thus for solutions that have good asymptotics at zero. 
Moreover, for this we needed solutions with imaginary energy (or action). 
Exactly like for the single DBI scalar whose 
solutions we described analytically in section 3, these are actually solutions 
to the action with $\beta^2<0$, analytically continued through their horizons. 

Indeed, we see now explicitly that in the $\beta^2<0$ case, any solution 
with good asymptotics at infinity will have a horizon at some $r_1$, exactly 
like in the case of the catenoids studied in section 3. As mentioned, if we 
analytically continue these solutions past their horizons we obtain solutions 
for the $\beta^2>0$ case with imaginary energy. 

We will see what happens to the temperature of the horizon with respect 
to the calculation in the previous section. We take a more general case of 
a Lagrangean of the type $\sqrt{f(\Phi)+(\vec{\nabla}\Phi)^2-\dot{\Phi}^2}$
such that near the horizon $r_1$ we have $\Phi ' \simeq \sqrt{f(\Phi)r_1/(2
(r-r_1))}$, as is the case here. Then we can easily calculate the fluctuation 
equation and find that near the horizon it looks the same way as in 
section 5, with 
$\sqrt{1+(\vec{\nabla}\Phi)^2}$ replaced by 
$\sqrt{f(\Phi)+(\vec{\nabla}\Phi)^2}$, but now we also get a ``mass term'' 
for the fluctuation, with 
\be
m^2= \frac{f''(\Phi)}{2}-\frac{(f'(\Phi))^2}{4(f(\Phi)+(\vec{\nabla}\Phi)^2)}
+\sqrt{f(\Phi)+(\vec{\nabla}\Phi)^2} \vec{\nabla}[\frac{\vec{\nabla}\Phi}{2
(f(\Phi)+(\vec{\nabla}\Phi)^2)^{3/2}}f'(\Phi)]
\ee
Then near the horizon 
\be
v^2/c^2= \frac{\Phi'^2}{f(\Phi)+\Phi '^2}\simeq \frac{r_1}{4(r-r_1)+r_1}
\ee
and we get the same result for the temperature as in the single scalar case, 
namely $T\propto (r_1)^{-1}$ (up to an infinite factor of the energy 
density), just that $r_1$ is now different. 

In conclusion, for $\beta^2>0$ (Pavlovskii's case) we have a topological 
solution, but no solutions with horizons, whereas for $\beta^2<0$ all 
solutions with good asymptotics at infinity have horizons, and there are 
no topological solutions. But we know from physical considerations that 
we would like an effective field theory for QCD that gives both.

\section{Proposed effective DBI action for QCD and ``SkyrBIon'' solution
for nucleons}

{\bf QCD effective action}

In fact, the action for a probe U(1) D-brane in the gravity dual geometry,
representing the dynamics of the IR cut-off of the geometry, 
has both a scalar and an electric potential, thus having both signs in 
the square root. As we saw in section 3, this action admits both BIon 
solutions, that are localized lumps of electric field analogous to the 
Skyrme-like soliton in the previous section, and catenoid solutions, 
that have thermal horizons, whose temperature we have calculated in section 5.
The BIons have arbitrary charges (given by their delta function
source), but in the quantum theory we expect those to be quantized. Their 
analog for Pavlovskii's SU(2) theory, the topological soliton, has already 
quantized charges, due to topology. 

The fact that the IR behaviour of the gravity dual can be encoded in a D-brane 
probe with a U(1) gauge field on it is not a priori obvious. However, this 
procedure has been used before to calculate meson masses. 

The usual procedure for having topological solitons of Skyrme type representing
the nucleons is to write a nonlinear extension of the sigma model action 
for the SU(2) pions. But then as we saw there seems to be impossible to 
have solutions with horizons as well, as we know we should, from the gravity 
dual picture. An action similar to the one of the U(1) D-brane probe would
give the correct physics, but why? We have neglected an important fact, namely 
that the action we are interested in has to be valid at energies beyond 
$\Lambda_{QCD}$. 

We know that in the analysis of QCD confinement one doesn't hope that the 
pions are the only effective degrees of freedom. They should be the only 
relevant degrees of freedom only at large enough distances (comparable with, 
or larger than $1/m_{\pi}$). 't Hooft  \cite{tho} proposed an idea that 
is now the standard way to understand confinement, the ``dual superconductor''.
Out of the SU(3) QCD action one, one uses a partial gauge condition that 
 keeps only the abelian projection, to the 
maximal abelian subgroup, $U(1)\times U(1)$. The normal vacuum of QCD is then 
a ``dual superconductor'' background: after a duality transformation
on the abelian subgroup $U(1)\times U(1)$
one is in a superconductor background, where the dual gauge field (monopole
field in QCD) is in the broken (massive) phase, due to interaction with a 
dual Higgs field, corresponding to a monopole-antimonopole pair condensate 
of QCD.  This dual Higgs field is of course also massive in the confining 
vacuum. 

Then a quark-antiquark pair of QCD (forming a meson)
will serve as endpoints for tubes of 
confined electric flux, as in a dual type II superconductor, giving a 
linear potential. In the dual 
abelian theory, monopole-antimonopole pairs (solitons of that $U(1)\times U(1)$
theory) will have between them magnetic field flux lines confined to tubes. 
In this dual theory, the gauge field will be nonzero inside the flux tube and 
zero outside (where it is massive), and the Higgs field will be zero at the 
core of the flux tube and go to the symmetry-breaking value on the outside.
Nucleons will have three quarks, thus are harder to explain, and one has to 
invoke some spherically-symmetric version of the meson mechanism. 
Also, in this picture the quarks are just sources, but if they are fundamental,
as in real QCD, they introduce extra light degrees of freedom, the SU(2) 
pions (in fact, the lightest). Also, the confinement picture changes, as 
quark-antiquark pairs can break off an electric  flux line. 

Seiberg and Witten \cite{sw} provided the first example of such a mechanism, 
for the SU(2) ${\cal N}=2$ SYM theory broken to the confining ${\cal N}=1$ 
SYM by a soft-breaking susy mass term $m Tr\Phi^2$.  The low energy effective 
action for the SU(2) ${\cal N}=2$ SYM theory is an abelian (U(1)) 
${\cal N}=2$ susy theory. At the point on the moduli space where monopoles 
become massless, one has to make a duality transformation and go to a 
description in terms of monopoles. When one adds the susy-breaking mass 
term, one is driven to the (former moduli space) point where monopoles are 
massless, and one finds that the dual Higgs field (monopole-antimonopole 
field) condenses, making the dual photon massive. This is the picture 
expected from the 't Hooft idea, just that it is valid only at low 
energies (on the moduli space). Seiberg and Witten calculate exactly only 
the low energy theory, given by $\int d^4 x d^4 \theta {\cal F}(\Psi)$, 
but higher order corrections (coming from terms like $\int d^4 x d^4 \theta
d^4\bar{\theta} {\cal H}(\Psi, \bar{\Psi})$, for instance) are not calculated.

So for the effective theory of QCD in the confining vacuum we expect 
to have a dual abelian (probably $U(1)\times U(1)$) theory for a gauge field 
interacting with a Higgs that gives the photons mass, and also the SU(2) 
pions (almost Goldstone bosons for the chiral symmetry introduced with 
the light fundamental quarks). The low energy $U(1)\times U(1)$ gauge theory
is probably complicated (as seen in the Seiberg-Witten example), and the 
low energy SU(2) pion action is the usual nonlinear sigma model action
\be
S=\int d^4 x \frac{f_{\pi}^2}{4} [Tr L_{\mu}L^{\mu}+ m_{\pi}^2
Tr (U+ U^+-2)]
\label{nlsm}
\ee
where as before $L_{\mu}=U^{-1} \partial_{\mu} U$. 
At high energies we don't have much information, except from the gravity 
dual theory. The action for a single scalar pion was of DBI type, as it 
corresponded to the position of a brane (IR cut-off) in an extra dimension. 
We know that the QCD pions are an SU(2) triplet, probably corresponding 
in the dual gravity theory to fluctuations of the IR cut-off (physical brane)
in an SU(2)-valued set of directions, like on an $S^3$. Thus we want to have 
an SU(2) generalization of the pion DBI action that reduces to (\ref{nlsm})
at low energies. 

For the dual gauge degrees of freedom, we will take a single U(1) (probably 
the correct theory involves both U(1)'s, but we take the minimal assumption).
We will also assume that for the problems at hand, the dual Higgs will only 
give mass to the dual photon. If the dual superconductor is correct in all 
details, the Higgs should change from zero inside the nucleons to the 
symmetry breaking value outside. However, from the gravity dual point of 
view an action for the Higgs would imply making assumptions about the actual 
model for gravity dual of QCD. Since we don't have such a model, we don't have
any idea how such an action would look like. The minimal assumption then is 
to hope that the dynamics of the Higgs is irrelevant for the nucleons, 
and either it is approximately constant, or
at the surface of the nucleon, where the Higgs would vary, the pion 
field is more important than the Higgs. In that case, we could just replace
the Higgs with a mass term for the U(1) gauge field. 
But what would be the high energy behaviour of the U(1) gauge action? 
From the dual picture, if we can think of the U(1) as the gauge field on a 
probe D-brane, the action should also be of DBI type. It is also hard to 
see what else could it be, as the DBI action is the unique 
nonlinear correction of 
the Maxwell theory that is both causal and has only one characteristic 
surface (generically, there are two) \cite{pleb}.

However, that can't be the gauge action for arbitrary energies either, we 
expect that above some energy the simplest description of the gauge fields 
should be in terms of the original gluons. Thus this action should be 
valid only for U(1) energies smaller than some scale $\bar{M}$, above which 
we cannot say anything. The simplest assumption that we will use afterwards, 
is that effectively $|\nabla \phi|\leq \bar{M}$, i.e. after it reaches 
$\bar{M}$ it stays fixed at that value, but in reality it should be a 
more complicated condition. The gravity dual picture suggests 
something similar, as we will discuss in more detail later. Indeed, we have 
seen that at $\hat{E}_R$ in the gravity dual the created black holes start 
feeling the curvature of space. Thus at this energy gravity 
(dual to the gauge fields) starts feeling new terms in the action, in 
agreement with having a relation between $\bar{M}$ and $\hat{E}_R$.

Thus we propose the action (at zero magnetic field)
\bea
&& S= \frac{f_{\pi}^2M_1^2}{2}\int d^4x  \nonumber\\&&
Tr[\sqrt{(1-\frac{\vec{E}^2}{M_2^2})
(1-\frac{L_i^2}{M_1^2})
- \frac{(E_i L_i)^2}{M_1^2M_2^2}+\frac{L_0^2}{M_1^2}+\frac{m_{\pi}^2}{M_1^2}
 (U+ U^+-2)+M_A^2 \phi^2}-1]\label{proposed}
\eea
where $\vec{E}=\vec{\nabla \phi}$ and $M_A$ is the gauge field mass, 
and $\vec{E}_{max}=\bar{M}$. 

The most conservative ansatz for the mass scales is $M_1=M_2=M_A=
\Lambda_{QCD}$, but they need to be only of the same order of magnitude 
from physical considerations. Most of the time we will assume this 
ansatz to be true however. 

A priori there should be some Chern-Simons term in the action as well, that 
reduces to the one in the original Skyrme model \cite{witten}, and since 
also the D-brane action in the gravity dual will have in general such a term. 
But we know that it will not contribute for spherically symmetric solutions
(which we are interested in), so we will ignore it in the following.

{\bf D-brane action toy model}

To understand the issues better, let us look at the simpler model of the 
single scalar pion, described by the U(1) D-brane action. This is the 
case analyzed in the gravity dual, with the D-brane scalar being the position 
of the IR cut-off (brane) in AdS space. Moreover, we analyzed its solutions 
in section 3, so now we can interpret them. In section 3 we looked at the 
massless action, which had as solutions the BIon, the catenoid, the BPS BIon 
and general solutions interpolating between these three. 

In the gravity dual, the cleanest case is the case when the scalar 
is also absent (pure gravity,
the IR cut-off is fixed, i.e. the radion has a very large mass).
That would correspond to putting the DBI scalar to zero. Then the BIon 
should stand for a glueball-type solution (excitation of the 
theory). The constant C in the BIon solution can be taken as electric 
charge (we identify it as such by the behaviour at infinity), thus
is quantized, as it should. But if the gauge field is massive, as we 
have argued it has to be (being in a dual superconductor background), 
we will not measure an electric charge at infinity.
However we also know that by colliding two glueballs we create 
black holes in the gravity dual, and eventually these black holes live 
on the IR cut-off, being effectively 4 dimensional. Thus eventually
(in the Froissart regime) in the 4 dimensional 
collision of two field theory
glueballs at very high energy we also expect to create a 
fireball type solution with a horizon. But we saw that the pure electric 
action (no scalar) has no solutions with horizons. It must follow that 
the description of the pure Yang Mills case at large energies cannot 
be in terms of the pure U(1) DBI action, but must include a true scalar
(for the description of a scalar glueball field), which scalar can generate 
solutions with horizons. This scalar should be relevant 
above some energy scale $\hat{E}_F$ (the unknown 
Froissart energy scale for glueballs). It could be for instance that the Higgs
field giving mass to the U(1) (which we ignored in this analysis) has also 
a DBI-like kinetic term with energy scale $\hat{E}_F$. 
But there is one more modification that one expects to the U(1) DBI action:
at energies above the soft Pomeron scale $\hat{E}_R$, in the gravity 
dual the created black holes start feeling the curvature of the space, thus 
the effective action for U(1) should get new terms. They will modify the 
value of $\phi $ only very slightly, but enough to ensure that 
for instance the energy density at the center of the BIon doesn't diverge 
anymore, but stays finite (the energy density is proportional to $1/\sqrt{
1-\phi '^2}$).

In the case that we have a nonzero 
single scalar pion (in the gravity dual the radion 
has a small mass), the BIon would again correspond to a glueball solution. 
Now a BIon-like general solution with $C>\bar{C}$ will be also a type 
of dressed glueball, and by colliding two such solutions
(at least for two solutions with opposite charge C so that the result 
of the collision is neutral) one should create a scalar pion fireball, with 
a horizon. Such a solution is the catenoid, and catenoid-like solutions 
with $\bar{C}>C$ if the total charge of the colliding objects is nonzero. 
But now there should be not only glueballs in the theory, but also 
hadrons, which have a scalar pion profile and quantized scalar charge
(toy version of the baryon charge). 
We need then $C\geq \bar{C}$ to avoid horizons, 
and if both have unit values, $C=\bar{C}=1$, 
which would imply the BPS BIon is such a hadron. However, we don't want 
for hadrons solutions with diverging scalar at zero, so we can either consider 
$C>\bar{C}$ and find a solution with a given finite value of the scalar
at the origin
$X(0)=X_0$ (which specifies $\bar{C}$ if C=1), or consider that the action 
has higher order corrections that limit the value of the derivative
$X'(0)$, such 
that the BPS BIon becomes also finite. In this toy model, the baryon charge 
is not well defined, but we will see that in the real case we will have a 
combination of the two proposed solutions, namely we will have a given value 
at zero $X(0)$ and the scalar derivative $X'(0)$ will be limited in value.

{\bf Solutions and SkyrBIon as nucleon}

We will first neglect the mass terms ($m_{\pi}^2$ and $m_A^2$ terms) in 
(\ref{proposed}). Note first that for $\vec{E}=0$ we get the Pavlovskii 
action with $\beta^2<0$, case we analyzed in section 6, and for U=1 we 
get the original action of Born and Infeld. Since the fields appear 
quadratically in the action, the above truncations are consistent, and 
we can embed any solutions we found in those cases in the action 
(\ref{proposed}) with $m_{\pi}=M_A=0$.  

We can  easily check that
$(n_i L_i)^2=-F'^2$, thus for a spherically symmetric solution, for which 
$E_i=-\partial_i \phi(r)\sim n_i$, $(E_iL_i)^2=-\phi '^2 F'^2$ and the \
action on the solution becomes
\bea
&&S=-\int dt \; r^2 dr [\sqrt{(1-\phi '^2 )(1+F'^2 +2\frac{\sin ^2 F}{r^2})
+\phi '^2 F'^2}-1]\nonumber\\
&& =-\int dt \; r^2 dr [\sqrt{(1-\phi '^2 )(1 +2\frac{\sin ^2 F}{r^2})
+F'^2}-1]
\eea
with equations of motion 
\be
[\frac{r^2 \phi ' + 2 \phi ' \sin ^2 F}{R}]'=0;\;\;\;
[\frac{r^2 F'}{R}]'=\frac{(1-\phi '^2) \sin 2 F}{R}
\label{mode}
\ee
where R is the square root in the action. 

The first equation can be easily integrated with a constant, giving
\bea
&& \phi ' = \frac{C}{\sqrt{C^2 +r^4(1+2\frac{sin^2 F }{r^2})}}
\sqrt{1+\frac{{F'}^2}{1+2\frac{sin^2 F }{r^2}}}\nonumber\\&&
R= \sqrt{\frac{r^4(1+2\frac{sin^2 F }{r^2})}{C^2+r^4(1+2\frac{sin^2 F }{r^2})
}}\sqrt{1+{F'}^2+2\frac{sin^2 F }{r^2}}
\label{phisol}
\eea

Then the second equation is (\ref{eom}) (for $\beta^2=-1/2$) with a modified 
right hand side, i.e.
\bea
&&(r^2+ 2\sin ^2 F )F''+ (2r F' -\sin 2F) \nonumber\\
&& +2 (r F'^3-F'^2 \sin 2F + \frac{3}{r} F' \sin ^2 F
-\frac{1}{r^2} \sin 2F \sin ^2 F)=\nonumber\\
&&= \frac{1+{F'}^2+2 \sin^2 F /r^2}{1+2 \sin ^2 F /r^2}\frac{C^2}{
C^2+ r^4 (1+2 \sin ^2 F /r^2)}\times \nonumber\\ && \times 
[2 r F' - \sin 2 F + 2 \sin ^2 F
(\frac{F'}{r}-\frac{\sin 2 F}{r^2})]
\label{modeom}
\eea

We can now find the behaviour of solutions at $r=0$ and 
$r=\infty$. Let's first assume that $F(0)=N\pi$ is a good vacuum at $r=0$
and expand around it. Defining $\bar{\phi}_0=\bar{\phi}(0)=
\phi '(0)$ and $ \bar{F}_0=\bar{F}(0)=F'(0)$, we get from (\ref{phisol}) that
\be
\bar{\phi}_0^2 =\frac{1+ 3\bar{F}_0^2}{(1+2\bar{F}_0^2)}
\ee
Note that this is the condition that R, the square root in the action, 
is zero at r=0, the same condition we obtained for the U(1) D-brane action 
in section 3, in particular for the BIon solution. Thus as for the DBI action
studied in section 3, $\bar{\phi}_0=1, \bar{F}_0=0$ is a solution, and is 
actually the same solution, {\bf the BIon}, since for F=0 we have the same 
action, and this is a consistent truncation of the original action. 

However, when we try to satisfy (\ref{modeom}) at nonzero F, we see that it 
is impossible. A Taylor expansion doesn't work, as the ${\cal O}(r)$ terms 
on the left hand side of (\ref{modeom}) cancel, whereas on the right hand 
side we get 
\be
\frac{1+3\bar{F}_0^2}{1+2\bar{F}_0^2}[-2r \bar{F}_0^3]
\ee
so that would imply $\bar{F}_0=0$. But then we would get 
\be
(r^2 F'')(0)=-2(r {F'}^3)(0)
\ee
which doesn't have any solution for any nonzero coefficient in the Taylor 
expansion. A singular behaviour of the type $F-N\pi = a r^{\alpha}$, with 
$0<\alpha<1$ doesn't work either, as it will give the equation 
\be
 a^3r^{3\alpha -2}\alpha (\alpha -1+i\sqrt{3})(\alpha-1-i\sqrt{3})=0
\ee
with no real solution between 0 and 1. The other possible singular behaviour, 
$F'= a \ln r, F-N\pi\sim ar(\ln r -1)$ doesn't give a solution either, 
the left hand side being of the order $r \ln ^2 r$ and the right hand side 
of the order $r \ln^3 r$. 

So if we would only consider (\ref{phisol}) we would obtain that, 
 as in section 3, $\bar{\phi}
_0=1\Rightarrow \bar{F}_0=0$, but now moreover $3/2\geq
\bar{\phi}_0^2\geq 1$, and $\bar{\phi}_0$ increases monotonically with
$\bar{F}_0$. Based on the toy model of section 3, we would expect to 
find solutions corresponding to all these
values, however as we saw, (\ref{modeom})
implies that in fact that all solutions must blow up before reaching r=0, 
as in the single scalar case! (Even in the pure pion case, at $\phi=0$, we 
had solutions that went to $F=N\pi$, though they had the wrong asymptotics 
at infinity, but now even those are excluded).

For the behaviour at
 $r=\infty$, we find as before, that the noninteracting asymptotics is 
not modified: Putting (and assuming $F(\infty )=0$)
\be
\bar{\phi}\sim \frac{a}{r^n};\;\;\; \bar{F}\sim \frac{b}{r^m}
\label{infi}
\ee
we find $n=2, m=3$ as for the case where the scalars don't interact with 
each other. 

For $\phi=0$ we have as we mentioned Pavlovskii's action for $\beta^2<0$, 
for which we saw that any solution that has good asymptotics at infinity 
will have a horizon. 

For a general solution (at nonzero $\phi$) 
with a good behaviour at infinity for
F, let's understand the formation of the horizon. 
If the horizon would be only 
in F, i.e. $F'=\infty$, F finite at r finite, and  $\phi'
$ finite, we would get 
\be
F\simeq F_0+\sqrt{\frac{(r_1^2+ 2\sin^2 F(r_1))(1-\phi '^2)}{r_1}(r-r_1)}
\ee
which is equal at $\phi'=0$ to the condition we had for the Pavlovskii case.
However, in fact we can see from (\ref{phisol}) that if we have 
F finite, $F'$ infinite at r finite, we also have $\phi '$ infinite. 
Then we get 
\be
F\simeq F_0+\sqrt{\frac{(r_1^2+ 2\sin^2 F(r_1))}{r_1(1-a)}(r-r_1)}
\ee
where
\be
a=\frac{C^2}{C^2+r_1^4(1+2\sin^2 F(r_1) /r_1^2)}\frac{1+\sin^2 F(r_1)/r_1^2}{
1+2\sin^2 F(r_1)/r_1^2}<1
\ee
thus we always have a solution, as in the Pavlovskii case. 

We also deduce that there can be no horizons of this type at r=0, where 
$\phi '\leq \sqrt{3/2}$
(with F finite and F' infinite, since that would need $\bar{\phi}_0\leq 1$, 
which is excluded), as we have already argued.

\begin{figure}

\begin{center}

\includegraphics{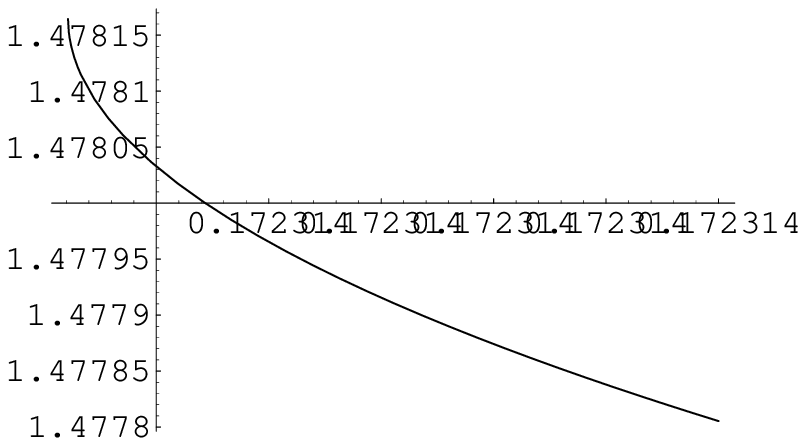}
\end{center}
\caption{Horizon detail}
\label{skyrbi}
\end{figure}

\begin{figure}

\begin{center}

\includegraphics{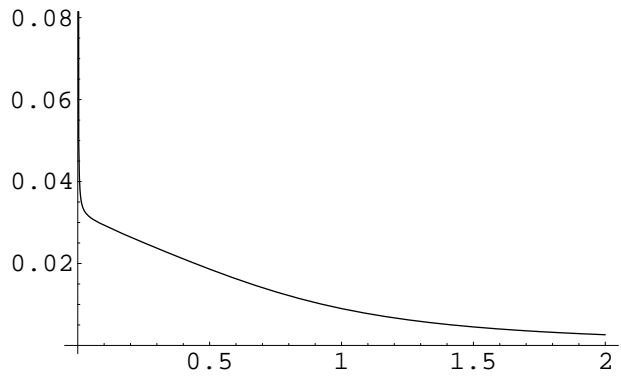}
\end{center}
\caption{Solution with larger b than the maximum}
\label{skyrbi2}
\end{figure}

\begin{figure}

\begin{center}

\includegraphics{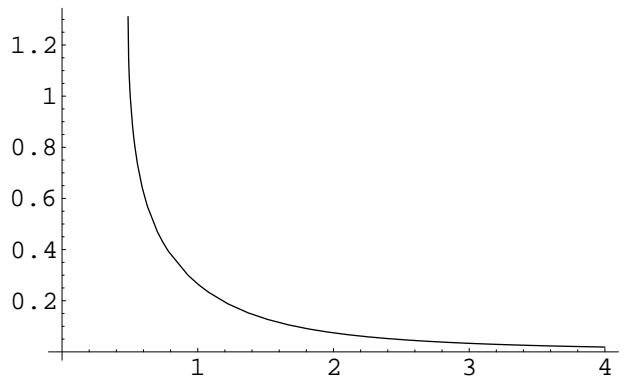}
\end{center}
\caption{Solution with smaller b than the maximum}
\label{skyrbi3}
\end{figure}

Finally, we have done an extensive numerical solution search for the 
equation (\ref{modeom}) confirming that there are no solutions that go to 
$F=N \pi$ at zero, and the solutions with good asymptotics at infinity 
have horizons at finite r. As the parameter b in the asymptotics at infinity 
of F in (\ref{infi}) is decreased, the position of the horizon $r_1$ 
decreases, and F increases until a maximum around $F\simeq 1.478$ for
$b\simeq 0.29$ and  $r_1\simeq 0.172$ (all this choosing C=1; 
see the horizon detail of this solution in fig.\ref{skyrbi}).
An example of horizon solution with $b=0.6, r_1\simeq 0.48$ and $F_m\simeq
1.31$ is given in fig.\ref{skyrbi2} and an example with $b=0.02, r_1\simeq 
0.001$ and $ F_m\simeq 0.28$ us given in fig.\ref{skyrbi3}, both of them 
having C=1. As one increases C, the maximum of F increases as well: for 
instance at C=3, for $b\simeq 1.8$ we have $F\simeq 2.18$ at $r_1\simeq 0.36$.

The conclusion is that this action (\ref{proposed}) does not have 
topological solutions, contrary to the intuition that we gained in the 
single pion case, the toy model D-brane action. However, we have to remember 
that we said the action (\ref{proposed}) is only valid up to $|\vec{E}_{max}|
=\bar{M}$. At F=0 there is no problem, as then $|\vec{E}|$ is bounded by 
$M_2$, but in the presence of diverging $F'$, $|\vec{E}|$ 
would diverge as well.

As we mentioned, the simplest assumption is to say $|\vec{E}|$ 
becomes constant 
and equal to $\bar{M}$ (in reality it could increase at a slower rate and 
saturate at a higher energy value). But by the relation between $F'$ and 
$\phi'$, $F'$ will be forced to saturate as well. Indeed, then the first 
equation in (\ref{mode}) can't be easily integrated to give (\ref{phisol})
anymore, but the modified second equation can be seen to still imply that 
F' is infinite only if $\phi'$ is infinite. We observe from (\ref{phisol}), 
valid before saturation, that the closer we are to r=0 when $\phi'$ saturates, 
the larger the saturation value of $F'$. 

Finally, this means that we will nevertheless have a topological solution, 
reaching $F=\pi$ because instead of the horizon in F at finite r of 
the action (\ref{proposed}), we can continue with $F'=F'_0>\bar{M}$ down 
to r=0. Numerically, this will work if $\bar{M}\sim 10$ (in units of 
$M_1=M_2$), since as we saw, for the maximum, $\pi -F_{max}\simeq 1.7$ and 
$r_1\simeq 0.17$ (and moreover, for larger $r_1$, $F_{max}$ decreases only 
slightly). Without a complete high energy action, we can't derive the exact
solution unfortunately. 

We will call this solution {\bf the skyrBIon} and it will represent a 
nucleon. 

Before we analyze it further, let us look at a point that may be of 
concern. Since we avoided the horizon in F at small r, why do we 
still have solutions with horizons at all (as we claimed, the solutions 
with horizons correspond to fireballs created in collisions)? 
The answer is that we have treated until now the massless case. But 
of course, we have nonzero masses for the pion and the gauge field. 
The mass of the gauge field should be around the first glueball mass, 
thus around $M_1,M_2$. The mass term will not affect the solutions at 
$r<M_A\sim M_1\sim 1$ (in our units), but at larger distances, $\phi$ 
will decreases exponentially to a negligible value. Thus if $F'$ diverges 
at $r>1$ it will not yet be coupled to the (very small) $\phi$ field
and thus $F'$ will not be bounded. Only at $r<1$ as for the skyrBIon solution
this can happen.

{\bf Interpretation}

Since the proposed skyrBIon solution will represent a nucleon, it has to 
have the right properties. The topological charge it carries, B, will be 
interpreted as baryon number, as usual. It can also be positive or negative,
corresponding to particles and antiparticles, as usual. 
It is not clear from what we 
derived whether there are solutions with higher baryon number, that will 
depend crucially on the high energy modification of the action. What seems 
certain is that there aren't any solutions with arbitrarily high
 baryon number, there will be a maximum B. That is so if the slope F' 
cannot increase indefinitely, since the point $r_1$ at which it starts 
saturating cannot be larger than $1/\Lambda_{QCD}$.
That is in accordance with the 
real world, where stable particles (spherically symmetric configurations) 
of arbitrarily high B don't exist, and solutions of $B>1$ if present could 
represent possible metastable particles. This situation is to be contrasted 
with, for instance, the case of SU(2) DBI pions of ``wrong'' sign  studied 
by Pavlovskii, where an arbitrarily high B was possible. We should mention 
that by contrast, nuclei (which have higher B), will not be spherically 
symmetric configurations, as is true in the real world. We will discuss 
them shortly. 

The skyrBIons have also a (quantized) U(1) electric charge. Due to the 
mass of the gauge field, it is not a charge detectable at infinity, 
but we will have skyrBIons of negative U(1) charge with the same $(\phi,F)$
profiles (the equations of motion are invariant under $\phi$ reflection), 
but different U(1) interactions. But that just accounts for the fact that 
the proton is different than the neutron, even once you take the 
electromagnetic interactions out! In other words, p-p vs. p-n scattering 
differs more than by the purely electromagnetic exchange diagrams. 
So the massive U(1) charge +1 and -1 particles with B=+1 are the p and the n. 

We also have the BIons in the theory, purely U(1) solutions
with an arbitrary quantized charge, 
but these are easily seen to be undestood as scalar glueball excitations, 
since the massive U(1) should be dual to the SU(3) gauge fields.

{\bf Nuclei as BIonic crystals}

Finally, how do we undestand the nuclei in this picture? It was understood 
for a long time that in Skyrme-like models, the nuclei should be bound states 
of nucleons forming sort of a crystalline structure and not by a higher B 
spherically symmetric soliton. This is indeed what we obtain, 
as we shall see, with the added 
bonus that it seems impossible to even have arbitrarily high B solitons anyway.

First, notice that BIon solutions of the massless BI action (U(1) gauge field)
can form crystals \cite{gibbons,hoppe} of the NaCl type, i.e. with positive 
and negative U(1) charges alternating in the crystal. This is maybe not so 
surprising, but one can write down an exact solution for this. If the
BIons in our theory represent glueballs as argued, it would be interesting to 
see what this fact could imply experimentally for glueballs. 
Of course, as we said, the 
gauge field is actually massive, so the BIon crystal would be modified in 
our theory anyway. 

More importantly, let's see what will happen for the skyrBIon solution. They 
come with massive U(1) charges +1 and -1, that we argued should correspond 
to the proton and neutron. At large distances ($r>M_A$) only the 
pion interaction is relevant, and it is an attractive interaction, being 
mediated by perturbative scalars. That is, a two skyrBIon solution
 at large distance will have a smaller energy than two individual skyrBIons. 

As we argued at the end of section 3, at small distances, as long as the 
energy is finite, we can figure out if the potential is attractive or 
repulsive from the behaviour as a function of charge. The energy of the 
static spherically symmetric 
system in the case of zero mass and no higher order corrections
is
\be
H=\int r^2 dr[\frac{1+{F'}^2+2\sin^2 F /r^2}{R}-1]
\ee
We see that only the $\sin^2 F/r^2$ terms stop the action from scaling 
as charge to the 3/2 power as in section 3, and if the radius
 where F' and $\phi'$ 
reach their maximum is not too small, this term  will not affect too much the 
scaling. We also see that before even reaching radii smaller than $1/M_A\sim
1/\Lambda_{QCD}$ and the U(1) field becoming relevant, the pion scalar field 
becomes nonlinear, and given the analysis in section 3 could already 
become repulsive.
We can thus say that the interaction will certainly
turn repulsive at some distance inside $1/\Lambda_{QCD}$, maybe even
before. 

Finally, we can say that once we reach the zone of saturation of the 
derivative of $\phi$ (at $r\simeq 
r_1$) the potential will be clearly repulsive.
Indeed, it is probably safe to assume that the energy density 
will be approximately constant 
($\rho_0$) inside it (for $r\leq r_1$). If F' is approximately constant inside
$r_1$, then $r_1$ is proportional to the topological charge B=N (at 
r=0, $F=N\pi$), thus the energy in the core will increase like $\rho_0
r_1^3\sim N^3$, thus two unit charge baryons will be strongly repulsed.  

Given this picture it is clear that there will be a minimum $r_C$
for the potential 
between two nucleons that is at least for radii
inside $1/\Lambda_{QCD}< 0.2 fm$ if not outside it (larger). Moreover, 
we talked about the interaction of two identical nucleons, but there will 
clearly be a diffence between the U(1) like charge and opposite charge 
interactions at least at distances of the order of $r_1$. 

All of the above features are in accordance with the real world, where a 
meson (pion) interaction potential works down to about 0.4fm after which 
the interaction potential  rises dramatically. 

The structure of nuclei is known to be well described by the
Bethe-Weizsacker formula for the binding energy of the nucleus
\be
B(A,Z)= a_VA-a_sA^{2/3}-a_C\frac{Z^2}{A^{1/3}}+a_P\delta A^{-3/4}-a_{sym}
\frac{(N-Z)^2}{A}
\ee
where $\delta =+1$ for (N,Z) even-even, -1 for odd-odd and 0 for even-odd 
or odd-even. 
The first three terms (volume term, surface tension
term and Coulomb interaction)
are known to be deduced from a liquid drop model, and the $a_P$ term is 
quantum in nature (the nucleons are fermions of spin 1/2). For the liquid 
drop model a nucleon-nucleon potential with a minimum is probably enough, and
that was qualitatively derived already. The last term in the binding energy 
can also be understood qualitatively in our model. As stated before, 
in the U(1) D-brane toy model, the BIons form NaCl-like crystals, with 
alternating +1 and -1 charges. Thus the minimum energy 
in this case is given by equal number of +1 charges and -1 charges. 
The same should be true if we add some (constant, positive) 
scalar charge to all BIons, such that we keep $C>\bar{C}$. 

It is then very likely that for skyrBIons also the minimum energy configuration
is given by equal numbers of U(1) charges +1 and -1, that we equated with 
n and p, thus for N=Z. Thus we expect indeed a term that drops with
$(N-Z)^2$ in the binding energy (the fact that it is $(N-Z)^2$ and not some 
other power is just because a perturbative U(1)
interaction is proportional to the 
charge squared).

\section{High energy scattering models}

In this section we will see what we can derive about the high energy scattering
of hadrons that are modelled by boosted field theory solutions. We 
will look both at the toy model solitons and at the skyrBIons. 

We will start with a few general observations.
If we take two boosted hadrons, represented by boosted solitons, and collide
them, we can follow Heisenberg's argument and find again the cross section:

If we have a coefficient of inelasticity (energy loss is ${\cal E}=\alpha 
\sqrt{s}$)
\be
\alpha \sim \frac{1}{b^n}
\ee
then, provided that the action is of DBI type, Heisenberg showed that 
$b_{max}$ is obtained when ${\cal E}=<E_0>\simeq m_{\pi}$ (for 
a linear action he gets $<E_0>\sim \gamma \sim \sqrt{s}$, thus gets a 
$b_{max}$ independent of s, so it's not relevant for the high energy 
behaviour of real hadrons). 

Then Heisenberg
 suggests that $\alpha$ should be proportional to the pion wavefunction
overlap, and the wavefunction is proportional to $e^{-mr}$, thus 
$\alpha \sim e^{-mb}$:
\be
\int d^3 r \psi_1(r)\psi_2(r)\sim \int d^3 r e^{-mr} e^{-m(b-r)}
\sim V e^{-mb}
\ee
after which one obtains $\sigma \sim ln^2 s$.
In our case though, if $\alpha \sim 1/b^p$, then we find $\sigma \sim s^{1/p}$.
But now if $\alpha$ is due to a wavefunction, pion or otherwise, and it
satisfies $\psi (r)\sim 1/r^n$, then 
\be
\alpha \sim \int  d^3 r \psi_1(r) \psi_2 (r) \sim \int d^3 r \frac{1}{r^n|
\vec{r}-\vec{b}|^n}\sim \frac{1}{b^{2n-3}}
\ee
thus $p=2n-3$,
contrasted with the case when $\alpha \sim \psi (b)$ which gives $\alpha\sim 
1/r^n$, thus $p=n$. Note that in Heisenberg's case there is no difference 
in between the two hypotheses about $\alpha$, nor if we also take derivatives
on the wavefunctions, in all cases we get $\alpha \sim e^{-mb}$. Now however, 
we get different results. Moreover, there is no good reason why the
 wavefunction overlap behaviour with s would dominate over the dynamics 
of the theory (in the Froissart regime, the overlap is exponential, thus
dominates). 

In the gravity dual case, we had a A-S shockwave perturbation in flat space 
or $AdS_{d+1}\times X_{\bar{d}}$, with  
\be
|\nabla \Psi(r)|\sim \frac{\sqrt{s}}{r^p}
\ee
and the horizon was situated where $|\nabla\Psi(r)|= 1$, giving 
$b_{max}\sim s^{1/(2p)}$, and since $\sigma \sim \pi b_{max}^2\sim s^{1/p}$, 
it is the same kind of analysis as we have now, if we identify $|\nabla
\Psi(r)|$ with 
${\cal E}(r)=\sqrt{s}
\alpha (r)$ (it would be also with $\sqrt{s}\psi(r)$ if $\alpha \sim \psi$). 

We should note that the quantity of interest is $\sqrt{s}\alpha (r)$, which
increases with $\sqrt{s}$. The pion field is a scalar, thus as we noted 
$\psi(r)$ (and thus $\alpha (r)$ when it is only due to the pion
wavefunctions) stays 
constant as one boosts. 

From the collision of high energy boosted solitons (BIons or skyrBIons) it is 
hard to derive the behaviour of $\alpha (r)$, but we can try to understand 
at least the energy regimes. All energy regimes of interest obey $\sqrt{s}
\gg \hat{M}\simeq \Lambda_{QCD}$ and then the soliton becomes a shockwave
(corresponding in the dual to $\sqrt{s}>M_P$, when particles become 
gravitational shockwaves). 

Now we will adress specific cases. First, when there are no pions at all
(in the gravity dual we only have gravity, the IR cut-off is non-dynamical), 
as we argued in the last section, we should have the massive U(1) field 
but in order to generate horizons we should have also a scalar field becoming
relevant at the Froissart scale $\hat{E}_F$. We will assume for generality 
that the mass of the gauge field, $M_A$, (as well as the mass of the 
new scalar field, $M_A'$, that should be of the same order) 
is smaller than the DBI scale 
of the gauge field, $M_1$. Then we have a region $1/M_A>r>1/M_1$ where we can 
still neglect the masses.  

We have seen in section 5 that if we can ignore 
the masses of the fields, the DBI action has a solution with a horizon 
and a temperature that goes like $T\sim M^{-1/3}$. In the collision of 
two glueballs we will create such a scalar field (Higgs?) fireball, thus 
there should be some evidence of thermalization in the final state (after 
the decay of the fireball). However, this object will not be dual to a 
black hole (but rather to black hole creation integrated over the 
5th dimension), so its analysis will be more complicated. 
It is hard to estimate the behaviour of $\alpha(r)$
and thus of $\sigma(s)$ at this point. It should depend on the massive
gauge field profile and on the scalar field (Higgs?) profile. 
As one continues increasing the energy, eventually the core of the BIon 
will become relevant. As we argued, when the scattering energy reaches 
 the ``soft Pomeron scale'' $\hat{E}_R$, in the gravity dual the produced 
black holes start feeling the curvature of the space, and correspondingly 
the U(1) DBI action should have high energy 
corrections that keep the energy density 
from diverging at the core of the BIon.  As one boosts the BIons
above $\hat{E}_R$, part of the interaction is due to the overlap of the 
BIon core with the tail of the second BIon, $\phi \sim \bar{C}_2/r$. Thus 
the high energy corrections to the BIon become important and the scattering 
cross section behaviour should be modified too.   
When $r_0\sim 1/M_A'$, thus 
when $\sqrt{s}\sim \hat{M}_P^4/{M_A'}^3$, one will still create a fireball, 
but the temperature will be different. The cross section behaviour however 
still depends on the dynamics of the theory, at least until an unknown 
energy scale $\hat{E}_F$. Note that $\hat{E}_F>\hat{E}_R>\hat{M}_P^4/{M_A'}^3$,
since from the gravity dual we know that 
$\hat{E}_R=\hat{M}_P^8/\Lambda_{QCD}^7=N^2 \Lambda_{QCD}
>\hat{M}_P^4/\Lambda_{QCD}^3$ and $\hat{E}_F>\hat{E}_R$. Thus the temperature 
of the fireball
will be modified before the onset of the modification of the $\sigma(s)$
behaviour (which should occur at $\hat{E}_R>\hat{M}_P^4/{M_A'}^3 $). 
Finally at $\hat{E}_F$ we will be in the Froissart 
regime. Indeed, a perturbative calculation for the horizon of the formed 
BIon at $r_0\gg 1/m_{\pi}$ (how much larger depends on what $\hat{E}_F$ is)
gives $\phi\sim \bar{C} e^{-M_A'r}/r\sim 1$, thus $b\sim  1/M_A' \ln(\bar{C}
M_A')$ and $\bar{C}\propto M^a$ (a=number). 

If we have a single scalar pion, i.e. the U(1) D-brane toy model (in the 
gravity dual the IR cut-off is dynamical now), we will also have a pion 
wavefunction. The fireball that will be created in the scattering will 
now be due to the pion field, which dominates over the second scalar, 
having larger wavefunction. 
The behaviour of $\alpha (r)$ and thus of $\sigma (s)$ until 
$r_0\sim 1/M_A'$ will still be dominated by glueball physics, but now the 
onset of Froissart behaviour is sometime after
$r_0\sim 1/m_{\pi}$, thus sometime after $\sqrt{s}\sim 
\hat{M}_P^4/m_{\pi}^3$, since the fireball is made up of the pion scalar 
field. In between $r_0\sim 1/M_A'$ and $r_0\sim 1/m_{\pi}$ the pion field
wavefunction predominates, but we still expect the dynamics of the theory 
(QCD) to dominate over the wavefunction overlap. Now we have the same 
onset of the soft Pomeron behaviour at $\hat{E}_R=\hat{M}_P^8/\Lambda_{QCD}^7=
N^2 \Lambda_{QCD}>\hat{M}_P^4/\Lambda_{QCD}^3$ (thus when $r_0$ is already 
$>1/M_A'$), but the Froissart behaviour will onset even later than before,
at $\hat{E}_F'$. After $\hat{E}_F'$ (when $r_0 \gg  1/m_{\pi}$)
we have Froissart behaviur in terms of $m_{\pi}$ instead of $M_A'$. 

Finally, if we have SU(2) pions, and the nucleons are modelled by the 
conjectured skyrBIon solutions, the overall picture is the same as for 
the single pion. The only difference is that now the skyrBIons will carry a
topological charge (baryon number), and again as before the energy density 
at the core of the solution will not diverge, due to higher order modifications
to the action. 

Another important issue to discuss is the issue of transparency of the 
scalar ``fireball''. We have seen that in all cases 
(above $\hat{M}_P$, even before the Froissart bound) we will produce 
metastable scalar field solutions with thermal horizons. We have also 
argued in section 5 that the horizon of the fireball acts like the horizon 
of a black hole with respect to scalar excitations, i.e. it will infinitely 
delay the exchange of information with the outside. 
This is exactly true if the temperature of the fireball is nonzero and 
finite, and if the temperature is infinite, is is true only for high 
energy modes (the phase and group velocities go like $1/\sqrt{k}$).
In \cite{nastase3} 
we have argued that the ``jet quenching'' observed at RHIC, which we argued 
should be in the Froissart regime, is nothing but the information paradox 
of black holes, i.e. information (quantum particles) coming in, and thermal 
radiation coming out. But we know that ``jet quenching'' is not absolute, 
there is information coming out, even in the RHIC (Froissart) regime, 
and certainly before that, so how do we reconcile this with the observation 
made that fireballs have information absorbing horizons? 

In the gravity dual, it is easy to understand this. One integrates over the 
fifth coordinate, i.e. the black hole produced in the bulk  has 
quantum fluctuations along the fifth dimension. As one increases the energy, 
the black hole becomes more classical and gets stuck on the IR brane. 
Thus particles thrown at the interacting region in 4d field theory can miss 
the black hole in the fifth dimension due to quantum fluctuations, but that 
becomes incresingly rare as the energy of the produced black hole
increases. But how do we understand 
this phenomenon in field theory? 

First off, we saw that for the catenoid (classical, static, spherically 
symmetric solution) the temperature is infinite and then the velocities
at the horizon go like $1/\sqrt{k}$, thus the infinite information 
delay at the horizon is only approximate, and valid only for high energy modes.
Since the RHIC experiments deals with high energy probes anyway (jets, 
hence the name ``jet quenching''), this is exacly what is observed. 
But we argued that the temperature becomes finite, due either to 
time dependence, higher order corrections to the action or 
quantum effects, in which case the horizon should absorb all information.
But 
while it is true that the fireball horizon absorbs scalar field information, 
it is also essentially (i.e. even classically) unstable: one has to continue 
the scalar field solution inside the horizon, and the most likely possibility, 
the ``brane bubble'' depicted in fig.\ref{cont}, is unstable, and 
there seems to be no way to continue to a singularity behind the horizon.
Another symptom of the classical instability is the infinite temperature of 
the simple catenoid, for which we said a possible cure would be through 
time dependence. 
The fireball also has a horizon, so it is clearly quantum mechanically
unstable also due to the thermal radiation. But unlike the black hole, 
the absence of a singularity that would crush everything makes it unclear 
why the information would be destroyed in the first place. A scalar particle  
entering the metastable scalar fireball would linger at the horizon for a 
long time, but as the fireball horizon would dissappear it could presumably 
continue forward. If the time scale of the fireball existence is long enough, 
the particle could thermalize, giving rise to ``jet quenching''. 

Note therefore that the scalar fireball is a much cleaner example of the 
information paradox, meaning that quantum particles collide and thermal 
radiation comes out, in a purely quantum mechanical scattering. This 
underscores the fact believed in string theory, that the black hole really 
doesn't destroy information \cite{malda2}, and the information can be 
retrieved from the {\em almost} thermal radiation coming out. 
The scalar fireball production is also a first step towards finding a 
field theory formalism where the temperature is not introduced by 
hand, but appears due to the time evolution (a zero temperature system 
creates a finite temperature one through time evolution).

\section{Discussion and conclusions}

In this paper we have analyzed the nucleons (baryons) 
and their high energy scattering 
in the fixed t, high s regime from the effective field theory point of 
view, guided by the gravity dual description developped in 
\cite{kn,kntwo,knthree,nastase2,nastase3}. At rest, the nucleons are supposed 
to be described by the Skyrme picture, as topological solitons of the pion 
field, and at high enough energy, by colliding shockwaves of the pion field, 
according to Heisenberg's model \cite{heis} for the saturation of the 
Froissart bound. We wanted a description that can interpolate between these 
two cases, with the nucleons being solitons of some effective action 
for QCD involving 
the pion field that when boosted to high energies become the colliding 
shockwaves.

Both in the Heisenberg model and the gravity dual descriptions, the DBI 
D-brane action played a major role, so we studied its static solutions, the 
BIon, the catenoid, the BPS BIon and the solutions interpolating between them.
The D-brane action was a toy model for the effective action for QCD, with the 
solutions interpolating between the BIon and the BPS BIon standing for the 
nucleons, and the solutions interpolating between the BPS BIon and the catenoid
standing for fireballs created in the high energy collision of nucleons. 
Since at the energy scale $\hat{E}_R$, in the gravity dual the action is 
changed by the fact that the created black holes start feeling the curvature 
of space, it follows that at high enough energies the same should be true 
for the U(1) effective gauge field. Its action should be modified at 
high energies, thus ensuring for instance that the energy density at the 
center of the BIon doesn't diverge. 
We analyzed the boosted solutions and then the fireballs that will be created
in their collision. The ``dumb holes'' of Unruh 
\cite{unruh} are an example of field theory (specifically, 
hydrodynamics equations)
solutions that have thermally emitting ``horizons'' where $v=c$ and $dv/dr$ 
is finite, such that (in Unruh's case, when $\rho$ and $c$ are nonzero and 
finite at the horizon)
\be
T=\frac{1}{2\pi} \frac{dv}{dr}
\ee
They can be in fact mapped to the black hole solutions, hence their name.
We have shown that catenoids are also of the same type, we can also 
map the fluctuation equation to a black hole equation, and the surface 
where the scalar field diverges, i.e. where $X '\rightarrow \infty$, while
$X$ is finite, also has $v=c$ and $dv/dr$ is consistent with
nonzero temperature. We calculated that 
the temperature of the static spherically symmetric solution is actually 
infinite, but argued that the infinite prefactor should be regulated, either 
classically or quantum mechanically and then the temperature scales as
\be
T\propto\frac{1}{ r_0}\propto \frac{\hat{M}_P^{4/3}}{M^{1/3}}
\ee
where $r_0$ is the horizon position, $\hat{M}_P$ is the DBI scale, and 
$M$ is the mass of the solution. We similarly 
 gave a perturbative argument for the case
where the scalar field has a mass $m$, that the temperature of the modified 
catenoid will be asymptotically (for large mass M) proportional to $m$. 
The horizon of the catenoid can be probed with other types of fields, but 
scalar excitations propagate as in a black hole background, thus the
propagation of information, described by their characteristic surfaces, will 
be obstructed (infinitely time delayed for high energy modes, the 
phase and group velocities scale as $1/\sqrt{k}$)
as for a black hole horizon, even though it takes a photon a finite time to 
go to the horizon and back (since we are in Minkowski space). We analyzed 
the propagation of waves in the scalar background, and proved that if the 
background corresponds to a black hole of finite nonzero temperature, then 
light takes an infinite geodesic time to reach the black hole 
horizon, and correspondingly for the scalar, the phase and group velocities 
go to zero at the horizon for all modes, besides scaling as $1/\sqrt{k}$. 

In the real world, the pions transform under a global SU(2), giving the 
possibility of having topologically stable solutions, that one identifies 
with the nucleons in the Skyrme program. In \cite{pav} it was shown that 
a wrong-sign SU(2) generalization of the DBI action has Skyrme-like solitons.
We analyzed this action in detail and found that there are no solutions with 
horizons, only topological solutions. For the action with the correct sign, 
we have again analyzed in detail and found that
any solution with the correct asymptotics at infinity will have a horizon, and 
there are no topological solutions. The temperature of the horizons is now 
also $T\propto(r_1)^{-1}$, but the position of the horizon, $r_1$, is modified 
with respect to the single scalar case. 

The creation and decay of the scalar fireball was found to be a clean 
laboratory for the black hole information paradox, since the same situation 
seems to appear: in high energy quantum collisions one creates thermally 
radiating fireballs, apparently violating the unitarity of quantum mechanics.
But this shows that the problem of the information paradox is not due to
quantum gravity, but rather to the lack of a unifying field theory formalism
where finite temperature can appear in the process of purely quantum mechanical
scattering at high energies, as we know happens in the RHIC experiments. 
There is another argument why the information paradox shouldn't be 
related to quantum gravity at all. In \cite{kss} a bound for shear 
viscosity over entropy density was proposed, $\eta/ s\geq 1/(4\pi)$, and 
in \cite{bl} it was shown that the bound is saturated by black holes in 
gravity duals. But in RHIC collisions, $\eta / s$ is close to the limit 
value \cite{shuryak}, a fact that we used in \cite{nastase3} to support our
assertion that RHIC fireballs are gravity dual black holes living on the 
IR brane. The gravity dual saturation of the 
bound was derived using apparently quantum gravity arguments for the 
thermodynamics of black holes, yet the 
bound itself is independent of the Planck scale $M_{Pl}$ (this being the 
reason why it is possible to have the same bound saturated at RHIC). That 
would suggest that the thermodynamics of black holes might not have anything 
to do specifically with quantum gravity, but might just come out of 
usual field theory. 

In section 7,
we  observed that the correct effective description in the real (QCD)
vacuum is given by the ``dual superconductor'' picture of 't Hooft. The 
dual of the maximal abelian subgroup $U(1)\times U(1)$ of $SU(3)$ is in a 
type II superconductor phase, the dual U(1) gauge fields being massive 
due to a Higgs field, except 
for flux tubes between dual monopole-antimonopole pairs representing mesons, 
with baryons being some spherically symmetric version of this picture. 
Thus the dual effective picture a la Seiberg-Witten 
for the nucleons in the usual vacuum should 
involve the U(1) gauge fields, the Higgs making them massive and the scalar 
SU(2) pion fields. Assuming that we can neglect the dynamics of the Higgs 
and postulating the DBI high energy form of the action based on the gravity 
dual as well as  Heisenberg's model, we wrote down an effective action for 
QCD in the usual vacuum, as given in (\ref{proposed}). It is a DBI action 
for the U(1) massive gauge field coupled to SU(2) light pions. The action 
has the nonsingular BIon solution, however, as soon as we add a scalar 
perturbation at infinity, the solution will be singular  at a finite radius, 
having a ``horizon'', with finite scalar F, but infinite F'. As a result,
there doesn't seem to be any topological solution to the action, only 
solutions with horizons. However, as 
argued, the DBI action must be modified where the energy density of the 
U(1) gauge field diverges, and thus we argued that there will be a solution 
to the modified action, that we dubbed {\bf skyrBIon}.

We argued that the topological charge B is the baryon  charge, as usual, 
with B=+1 being baryons, B=-1 antibaryons and higher B solutions, if present, 
representing states unstable against decay into B=1 states
because of energy considerations. The objects of charge +1 and -1 under the 
massive U(1) should represent the n and the p
(the n and the p have different quark content, which will therefore have 
different gluon interactions). We argued that the potential 
between two nucleons should have a minimum at an $r_C$ of the order of 
$1/\Lambda_{QCD}\sim 0.2fm$. This feature is enough to guarantee that 
a nucleus can form from individual nucleons, in a liquid drop-like model.   
Given that BIons form NaCl-like crystals of
alternating charges, we argued that the nuclear energy  for skyrBIon 
nuclei should be minimized for N=Z, giving a qualitative agreement with 
the Bethe-Weizsacker formula. 

For the high energy (high s, fixed t) scattering of skyrBIons we could 
find qualitative agreement with the gravity dual picture, by identifying 
$|\nabla \Psi(r) |$ with ${\cal E}(r)=\sqrt{s}\alpha (r)$, but the 
particular small power laws obtained in the gravity dual cannot be 
deduced just from the effective DBI action. In the high energy scattering 
of two baryon-like solutions of the U(1) D-brane toy model one will create 
a catenoid=
scalar fireball of temperature that goes like $T\propto M^{-1/3}$, where M 
is the total energy of the collision. In this regime, the fireball does not 
behave yet like a black hole, except for having a  temperature.  
In the case of skyrBIon collisions, 
the created fireball solution will be similar.
We have argued for the existence of a ``soft Pomeron scale'' and a 
Froissart onset scale, as we derived from the gravity dual picture, but 
we can't calculate them, except to say that both should be larger than 
$\hat{M}_P^4/\Lambda_{QCD}^3= N\Lambda_{QCD}$. The Froissart regime is 
associated with production of 
a scalar fireball of size much larger than $1/m_{\pi}$, and the temperature 
of the fireball should be proportional to $m_{\pi}$ in this asymptotic regime, 
as argued in \cite{nastase3} from a gravity dual point of view. Then the 
produced scalar fireball is directly mapped to a gravity dual black hole 
living on the 4 dimensional IR brane. 

Thus as we advocated in \cite{nastase3}, we have shown that we can 
describe the gravity dual picture for high energy scattering
completely in terms of field theory, 
but we have seen the limitations of field theory in terms of calculability. 
The thermal property of black holes in the gravity dual is easily 
understood as the thermal property of ``horizons'' in the effective pion 
field. The colliding pion field shockwaves are seen as being just boosted 
versions of nucleons. The nucleons are described as solutions of an effective 
action, and lead to good qualitative features for the description of nuclei.

{\bf Acknowledgements} 
I would like to thank S.G. Rajeev for participation at the initial stages 
of this project, including for the idea of the project itself. 
I would also like to thank Antal Jevicki, Juan Maldacena
 and Jeff Murugan for useful discussions.
This research was  supported in part by DOE
grant DE-FE0291ER40688-Task A.

\end{document}